\documentclass[aps,prd,superscriptaddress,preprintnumbers,nofootinbib,11pt]{revtex4-1}

\usepackage{amsmath,amssymb}
\usepackage{graphicx}
\usepackage{color}
\usepackage{units}
\usepackage[hyperfootnotes=false,colorlinks,citecolor=blue]{hyperref}
\usepackage{slashed}

\newcommand{\beq}{\begin{equation}}
\newcommand{\eeq}{\end{equation}}
\newcommand{\bea}{\begin{eqnarray}}
\newcommand{\ena}{\end{eqnarray}}

\newcommand{\dd}{{\rm d}}
\newcommand{\MSun}{{\ifmmode{{\rm{M_{\odot}}}}\else{${\rm{M_{\odot}}}$}\fi}}

\begin{document}
%%%%%%%%%%%%%%%%%%%%%%%%%%%%%%%%%

\preprint{ULB-TH/24-01}
\title{Dark matter bound-state formation in the Sun
}

\author{Xiaoyong Chu}
\email{xiaoyong.chu@oeaw.ac.at }
\affiliation{Institute of High Energy Physics, Austrian Academy of Sciences, Nikolsdorfer Gasse 18, 1050 Vienna, Austria
}
\author{Raghuveer Garani}
\email{garani@fi.infn.it}
\affiliation{INFN Sezione di Firenze, Via G. Sansone 1, I-50019 Sesto Fiorentino, Italy} 
\author{Camilo Garc\'ia-Cely}
\email{camilo.garcia@ific.uv.es}
\affiliation{Instituto de F\'{i}sica Corpuscular (IFIC), Universitat de Val\`{e}ncia-CSIC, Parc Cient\'{i}fic UV, C/ Catedr\'{a}tico Jos\'{e} Beltr\'{a}n 2,
E-46980 Paterna, Spain}
\author{Thomas Hambye}
\email{thomas.hambye@ulb.be}
\affiliation{Service de Physique Th\'eorique  Universit\'e Libre de Bruxelles,
Boulevard du Triomphe, CP225, 1050 Brussels, Belgium}
\affiliation{CERN, Theoretical Physics Department, Geneva, Switzerland}

%%%%%%%%%%%%%%%%%%%%%%%%%%%%%%%%%%
%%%%%%%%%%%%%%%%%%%%%%%%%%%%%%%%%%

\begin{abstract}

The Sun may capture asymmetric dark matter (DM), which can subsequently form bound-states through the radiative emission of a sub-GeV scalar. This process enables generation of scalars without requiring DM annihilation. In addition to DM capture on nucleons, the DM-scalar coupling responsible for bound-state formation also induces capture from self-scatterings of ambient DM particles with DM particles already captured, as well as with DM bound-states formed in-situ within the Sun.  This scenario is studied in detail by solving Boltzmann equations numerically and analytically.  In particular, we take into consideration that the DM self-capture rates  require a  treatment beyond the conventional Born approximation. We show  that, thanks to DM scatterings on bound-states, the number of DM particles captured  increases exponentially, leading to enhanced emission of relativistic scalars through bound-state formation, whose final decay products could be observable. We explore phenomenological signatures with the example that the scalar mediator decays to neutrinos. We find that the  neutrino flux emitted can be comparable to atmospheric neutrino fluxes within the range of energies below one hundred MeV. Future facilities like Hyper-K, and direct DM detection experiments can further test such scenario.

\end{abstract}

%%%%%%%%%%%%%%%%%%%%%%%%%%%%%%%%%
%%%%%%%%%%%%%%%%%%%%%%%%%%%%%%%%%
\maketitle
{
  \hypersetup{linkcolor=black}
  \tableofcontents
}
%%%%%%%%%%%

\section{Introduction}

Wherever dark matter (DM) particles are numerous, it is possible that two or more of them form bound-states~\cite{March-Russell:2008klu,Pospelov:2008jd,Shepherd:2009sa,Laha:2013gva,vonHarling:2014kha,Petraki:2014uza,Petraki:2015hla,Pearce:2015zca,Laha:2015yoa,An:2015pva,An:2016gad,An:2016kie,Bi:2016gca,Kouvaris:2016ltf,Cirelli:2016rnw,Asadi:2016ybp,Petraki:2016cnz,Johnson:2016sjs,Mitridate:2017izz,Baldes:2017gzu}. This could occur if DM undergoes  attractive self-interactions mediated by a scalar or vector boson. In this case bound-state formation (BSF) can occur radiatively via the emission of this mediator, leading to an observable flux of particles. Indirect signals of BSF in the center of the Milky Way has been studied for DM with no particle-antiparticle asymmetry, as well as for asymmetric DM \cite{Pearce:2013ola,Baldes:2017gzu,Mahbubani:2019pij,Mahbubani:2020knq,Baldes:2020hwx}. In this work we consider the possibility that the BSF process occurs in the Sun and study the corresponding  DM indirect detection signals.

Symmetric DM accumulating in the Sun or other celestial bodies from its capture on nucleons and their corresponding indirect detection signals due to emitted meta-stable mediators have been studied in numerous works, see e.g.~\cite{Chen:2015uha, Feng:2016ijc, Arina:2017sng, Leane:2017vag, Lucente:2017jcp,Niblaeus:2019gjk, Cuoco:2019mlb,Mazziotta:2020foa, Dasgupta:2020dik, Bell:2021pyy, Garani:2021feo, DUNE:2021gbm, Leane:2021ihh, IceCube:2021xzo, Bell:2021esh,Zakeri:2021cur,Bose:2021cou,Maity:2023rez, DuttaBanik:2023yxj,Nguyen:2022zwb,Linden:2024uph}.
In these scenarios, the accumulation of DM particles in the Sun reaches a saturation point when the annihilation rate matches the capture rate. This is not the case of asymmetric DM scenarios as DM cannot annihilate. Interestingly, this absence of annihilation permits a greater accumulation of DM particles. However, it comes with the drawback of not generating any indirect detection signal. 
 
The BSF of asymmetric DM particles in the Sun from radiative emission of a light scalar has the interesting property of allowing both DM indirect detection and large accumulation of DM particles in the Sun.  A crucial aspect of this scenario is that the DM-scalar interaction that is needed for BSF inherently implies that DM capture results not only from interactions with nucleons but also from interactions with previously accumulated DM particles and DM bound-states (DMBS). This means that the capture rate is larger than that of the usual symmetric or asymmetric scenarios in which the capture only arises from DM-nucleon scatterings. This allows for a larger accumulation and an enhanced flux of emitted particles. To quantify this effect, it will be necessary to compute the rates for DM-DM and DM-DMBS scatterings, which --as we will show-- receive non-perturbative contributions which can be calculated in the semi-classical approximation.  To our knowledge this possibility that asymmetric DM particles form bound-states in the Sun and that DM particles are captured by scattering off DM bound-states has not been considered before.\footnote{Particle-antiparticle BSF inside the Sun for symmetric DM was considered in \cite{Kouvaris:2016ltf}.}

For concreteness, in this work we assume that the associated light scalars, once emitted when the bound-states form, decay into SM particles, in particular to neutrinos. As is well known, unlike other SM particles, low energy neutrinos can escape the Sun leading to observable signatures  even if the decay takes place inside the Sun. After solving the set of Boltzmann equations that describe the dynamics of DM accumulation in the Sun in section~\ref{sec:set-up},  we show in section~\ref{sec:neutrinoflux} that an observable neutrino flux  could arise   from the decays of the emitted mediator via the BSF of those accumulated DM.  
Finally, in section~\ref{sec:bounds} we discuss relevant constraints from both astrophysical observations and terrestrial experiments, such as DM self-interactions, BBN, CMB, and direct/indirect searches. 
%
%
%%%%%%%%%%%%%%%%%%%%%%%%%%%%%%%%%%%%%%%%%%
\section{Number evolution of DM particles}
\label{sec:set-up}

Before considering quantitatively any concrete model, this section introduces the basic relevant processes and related Boltzmann equations determining the number of DM particles captured in the Sun. The framework of our interest is based on an interaction between the DM particle and a lighter particle. For definiteness in the following we consider that DM is made of Dirac fermions $\chi$, with a  Yukawa coupling $g_s$ to a lighter scalar particle $\phi$ 
\begin{equation}
{\cal L}\supset -g_s \phi\bar{\chi}\chi \,.
\end{equation}
 To assume a scalar mediator is convenient because it induces an attractive interaction.\footnote{ Such interactions do not result in the formation of mini black hole in the Sun, i.e. the Chandrasekhar limit for fermions is not modified~\cite{Gresham:2018rqo,Garani:2022quc}.} 
 We also assume that DM is asymmetric, thus it does not annihilate in the present epoch. Nevertheless, in addition to the usual capture due to scattering on target nuclei ($C_\star$)~\cite{Press:1985ug,Griest:1986yu},  three additional terms come into play in the Boltzmann equation determining the DM particle number $N_\chi$, due to  DM-$\phi$ interactions. First, depending on the strength of self-interaction, DM particles could efficiently form  $\chi$-$\chi$ bound-states by emitting the particle mediating DM self-interactions. The rate of the latter is denoted by $A_{\rm bsf}$. Since the number of free particles is reduced by two per process, the term $- 2\cdot \frac{1}{2}A_{\rm bsf} N^2_\chi$ is introduced in the equation that describes the evolution of  $N_\chi$, where the second factor, $\frac{1}{2}$, counts for double-counting of identical initial states. In contrast to the case of annihilation, in this scenario DM particles are not lost by the system. In effect, there is a build-up of DM bound-states within the celestial body, whose number $N_{2\chi}$ is determined by a second Boltzmann equation that simply involves a $+\frac{1}{2}A_{\rm bsf} N^2_\chi$ term.  Second, there is a term coming from capture of Galactic DM on already-captured DM particles, whose rate is denoted by $C_\chi$. Finally, there is a term from capture of Galactic DM on the formed  bound-states, denoted by $C_{2\chi}$. The presence of both $C_\chi$ and $C_{2\chi}$ terms increase DM accretion. In summary, both populations evolve according to the following set of differential equations,
 \bea
\label{eq:boltz}
\frac{\dd N_{\chi}}{\dd t} &=& C_\star - A_{\rm bsf} N^2_\chi + C_\chi N_\chi + C_{2\chi} N_{2\chi}\,,\\ 
\label{eq:boltz2}
\frac{\dd N_{2\chi}}{\dd t} &=& \frac{1}{2} A_{\rm bsf} N^2_\chi~. 
\ena
The various rates $C_\star$, $C_\chi$ and $C_{2\chi}$ take into account the fact that the three kinds of DM capture occur within different spheres around the solar center. Here we have neglected self-capture via direct BSF between a galactic DM and a captured DM particle, as such inelastic scattering is typically much weaker than the elastic scattering with two DM initial states.  
{We further assume that the scattering between a galactic DM and a captured bound-state is also elastic, neglecting the possible formation of three-body bound-states or bound-state dissociation. }
Also, in these Boltzmann equations we do not write down explicitly additional terms coming from possible evaporation of the captured DM particles. We have checked that the DM mass above which evaporation is negligible is not significantly different than in the standard scenario (no-self interaction effects). The  value is found to be $m_{\rm evap}\approx 5$ GeV, which is about 50\% larger than that in the standard scenario~\cite{Garani:2021feo}. Therefore, in the following phenomenological discussion we consider only DM masses above $5$\,GeV.  
Once DM particles are captured they thermalize with the solar material, leading to efficient formation of bound-states.
We always assume that the thermalization happens quickly, and refer to   Appendix~\ref{ap:DMthermalization} for   detailed calculation of the thermalization process.  

The number of DM particles captured in any of the three ways cannot be larger than the corresponding geometric rates,  as the latter assume that all DM particles crossing the corresponding thermal spheres are captured. 
This is taken into account through the following conservative matching for the regimes of small optical depth (thin) to large (thick)\footnote{These equations are accurate if only one geometric rate can be saturated. This will be the case studied here, for which   the geometric capture rates on free DM ($C^{g}_\chi$) and on nucleons ($C^{g}_\star$) are never reached.}
\bea
C_\chi N_\chi &\rightarrow& C_\chi N_\chi \Theta(C^{g}_\chi - C_\chi N_\chi) + C^{g}_\chi \Theta(C_\chi N_\chi-C^{g}_\chi)~,\\ 
C_{2\chi} N_{2\chi} &\rightarrow & C_{2\chi} N_{2\chi} \Theta(C^{g}_{2\chi} - C_{2\chi} N_{2\chi})+ C^{g}_{2\chi} \Theta(C_{2\chi} N_{2\chi}-C^{g}_{2\chi})~. 
\ena
 where the  respective geometric rates on the DM and DMBS thermal spheres are denoted by $C^g_\chi$ and $C^g_{2\chi}$. These geometric rates determine the maximal possible capture rates, independently of the underlying particle model.

 Note that we do not consider bound-states containing more than two $\chi$ particles, under the assumption that there exist bottlenecks to form heavier bound-states, such as fast decays of $(3\chi) \to (2\chi) + \chi $ and $(4\chi) \to (2\chi) + (2\chi)$.  See e.g. \cite{Wise:2014ola, Hardy:2014mqa, Gresham:2017cvl,Gresham:2017zqi} for further discussions. In practice, if there is formation of many-body bound-states, each captured  DM particle may cause the emission of a few more mediator particles while thermal radii for heavier bound-states shrink by a factor of few. The total effect at most modifies our results mildly.\footnote{The presence of stable many-body bound-states, as well as self-capture via inelastic scattering, may reduce DM evaporation efficiency by quickly capturing free $\chi$ particles into  heavier bound-states. This can have qualitative consequences for DM candidates below GeV or those captured by Earth, which is left for a future study. }

We begin in the next subsection by determining the various rates. The reader seeking immediate understanding of their interplay in the Boltzmann equations can directly refer to section~\ref{sec:2solutions}.

%%%%%%%%%%%%%%%%%%%%%%%%%
\subsection{Determination of the various rates}

%%%%%%%%%%%%%%%%%%%%%%%%%
\paragraph{\underline{Thermal radius and geometric rates $C^g_\star$, $C^g_\chi$ and $C^g_{2\chi}$:}}  Once a DM particle has been captured in the Sun, or a DMBS has formed, these particles will thermalize with the SM material of the Sun and lie within different spheres of thermal radius $r_{\rm th}$. Noting that the mass of a bound-state is approximately twice that of a DM particle, these radii are obtained~\cite{Spergel:1984re,Press:1985ug}   by equating the average thermal energy, $3T_\odot/2 $,  to the  gravitational potential energy per particle, $2\pi G r^2_{\rm th} \rho_\odot  n m_\chi/3$,   as  
\begin{align}
\!\!r_{\rm th} \!= \!\left(\frac{9 T_{\odot}}{ 4 \pi G \rho_\odot n m_\chi}\right)^{\frac{1}{2}}\!\!= 0.03 R_{\odot}\!\! \left(\frac{T_{\odot}} {2.2 \, {\rm keV}}\frac{150 \, {\rm \frac{g}{cm^3}}}{\rho_{\odot}} \frac{\rm 10\,GeV}{n\, m_\chi}\right)^{\frac{1}{2}}\!\!,&&\!
\text{with $n=1(2)$ for DM (DMBS)}.
\end{align}
That is, the DM bound-state thermal radius is smaller than the DM thermal radius by a factor of $\sqrt{2}$. We take the solar core temperature, $T_{\odot}$, to be $2.2\,$keV, and the core mass density to be $\rho_\odot\sim $150\,g/cm$^3$. For example, the thermal radius of $3$\,GeV DM particles is about one-tenth of the Solar radius, $R_\odot \simeq 6.9\times 10^5$\,km~\cite{Asplund:2009fu}. In addition, we will assume throughout that the DM radial distribution is isothermal, with its temperature $T\approx T_{\odot}$~\cite{Press:1985ug,Garani:2017jcj}.

Upon taking into account the relative motion of the Sun with respect to the galactic DM halo, the geometric capture rate on nucleons is~\cite{Bottino:2002pd}
\begin{equation}
	C^g_\star \,\,{ =  \frac{\rho_\chi}{m_\chi}\, \pi R_\odot^2 \,\bar{v} }
 \simeq 
 {6 \times 10^{28}}\,{\unit{s}}^{-1}\left( \frac{\rho_\chi}{0.3\, \unit{GeV/cm^3}} \cdot \frac{10\,\unit{GeV}}{m_\chi} \right)\,.
	\label{eq:capturegeom}
\end{equation}
Here $\bar{v} $ is a factor with units of velocity, accounting for the relative motion of the Sun and the velocity distribution of DM, see e.g.~\cite{Garani:2017jcj}. Throughout this work we take the local DM density $\rho_\chi = 0.3$\,GeV/cm$^3$. Apparently, the geometric capture rate is proportional to the corresponding cross-sectional area. So this expression can be rescaled to obtain the geometric rate for DM self-capture on DM and on DMBS (i.e. when the mean free path of DM is smaller than the corresponding thermal sphere) by adding the corresponding factor  $(r_{\rm th}/R_\odot)^2$. This results in geometric self-capture rates which scale as $m^{-2}_\chi$, 
\bea
C^g_\chi &=& 5 \times 10^{25}\, {\rm s^{-1}} \left(\frac{10\,{\rm GeV}}{m_\chi} \right)^2\,, \quad \quad C^g_{2\chi} \approx 2.6\times 10^{25} \,{\rm s^{-1}} \left(\frac{10\,{\rm GeV}}{m_\chi} \right)^2\,.
\label{eq:selfcapgeom}
\ena

%%%%%%%%%%%%%%%%%%%%%%%%%%%%
\paragraph{\underline{Capture on nucleons rate $C_\star$}:} The capture rate of DM particles in the Sun has been well studied in the literature. It can be written as~\cite{Feng:2016ijc}
\bea \label{eq:cap-full}
C_\star &\approx&  \sum_i \int_0^{R_\odot} 4 \pi r^2 n_i(r) \dd r \int_0^\infty \dd u_\chi \, \left(\frac{\rho_\chi}{m_\chi}\right) \times \,  u_\chi \, \omega^2(r)f_{\odot}(u_\chi)  \int_{E_R^{min}}^{E_R^{max}} \frac{\dd \sigma_i}{\dd E_R} \,  \dd E_R \,, 
  \ena
where $f_{\odot}(u_\chi)$ is the normalized asymptotic DM velocity distribution far from the Sun in the solar frame. The values of radial number density distribution of each element $i$, denoted as $n_i(r)$ above,  are  adopted from the AGSS09 Solar model~\cite{Asplund:2009fu}. Together with the escape velocity at a distance $r$ away from the centre of the Sun, $v_e(r)$,  it provides the relative velocity of a DM particle when it scatters with the nucleus, $\omega(r) = \sqrt{u_\chi^2 +v_e(r)^2}$. The differential cross section, ${\dd \sigma_i}/{\dd E_R}$, encodes the energy dependence of elastic scattering between DM and nucleus $i$ in the non-relativistic limit. The recoil energy in the solar frame is given by $
 E_R = \mu_{\chi N}^2/m_N \times w^2 (1- \cos \theta_{CM})\,,$
 where $\mu_{\chi N}=m_\chi m_{N}/(m_\chi +m_{N})$ is the reduced mass of the DM-nucleus system. 
For the Solar capture to actually occur,  we need to make sure that the recoil energy lies within the range of  $E_R^{min} <E_R <E_R^{max}$, with 
 \beq
 E_R^{min} =\frac{1}{2} m_\chi u_\chi^2 ,\, \quad    E_R^{max} = \frac{2 \mu_{\chi N}^2 \omega^2}{m_N}\,.
 \eeq
For definiteness we assume that the dark sector communicates with the SM particles through the Higgs portal, and that the light mediator $\phi$ mixes with the Higgs boson with a mixing angle $\theta_\phi$. The differential cross section in the Born approximation is then
\beq
\frac{\dd \sigma}{\dd E_R} = \frac{g^2_s \cos^2 {\theta_\phi}  \sin^2 {\theta_\phi} }{2 \pi} \frac{m^2_N f^2_N}{v^2_H} \frac{m_N}{\omega^2(2 m_N E_R + m^2_\phi)^2} F^2\left(\frac{E_R}{Q_0} \right)~\,,
\label{Cstar}
\eeq
where $F(E_R/Q_0)$ is the Helm form factor and $f_N = f^N_u + f^N_d + f^N_s + 6/27 f_G$, which is an $\mathcal{O}(1)$ number (see e.g.~\cite{Lewin:1995rx, Duda:2006uk}). 
 
\begin{figure}[htb!]
     \centering
    \includegraphics[width =0.495\textwidth]{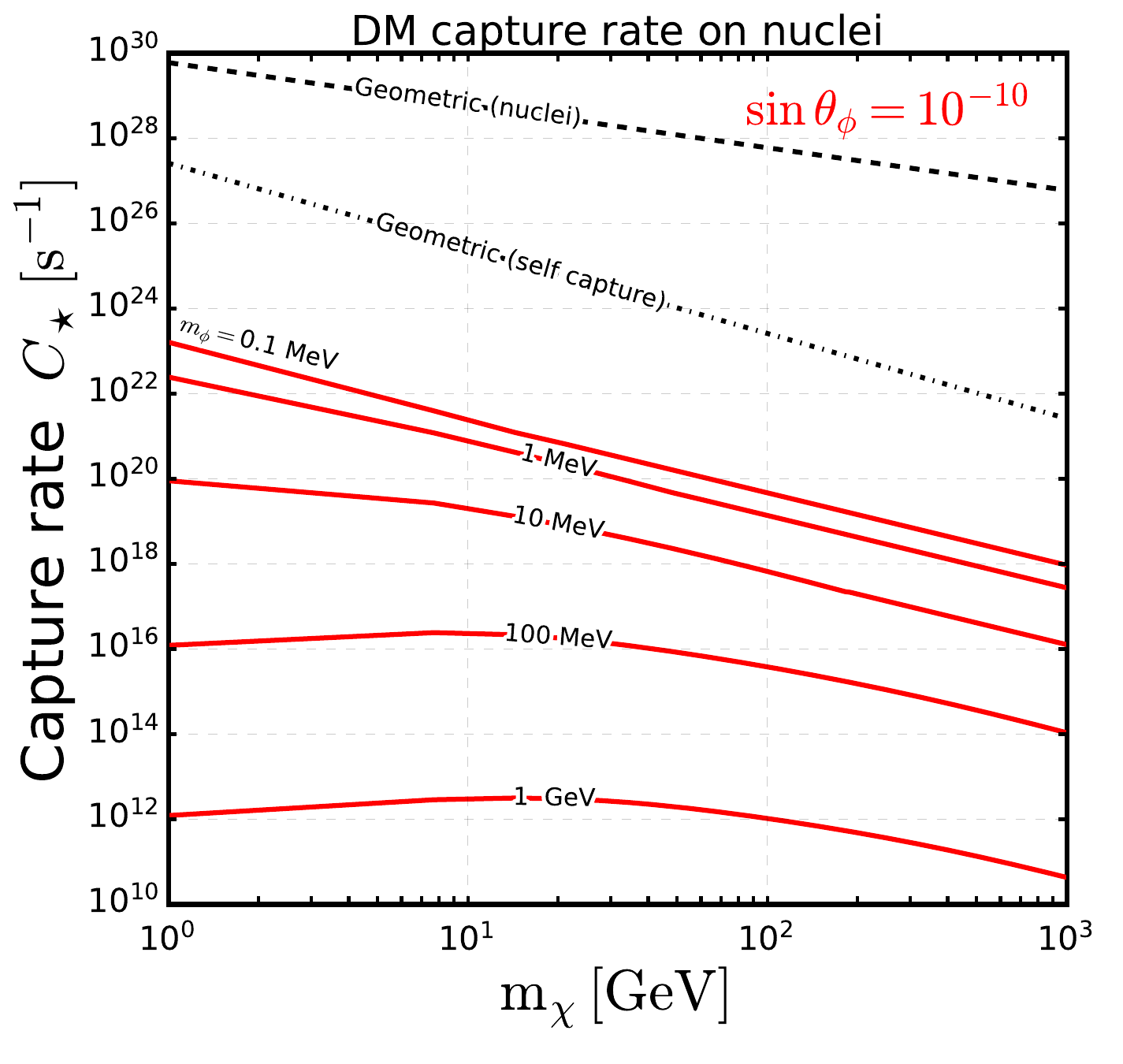} \includegraphics[width =0.495\textwidth]{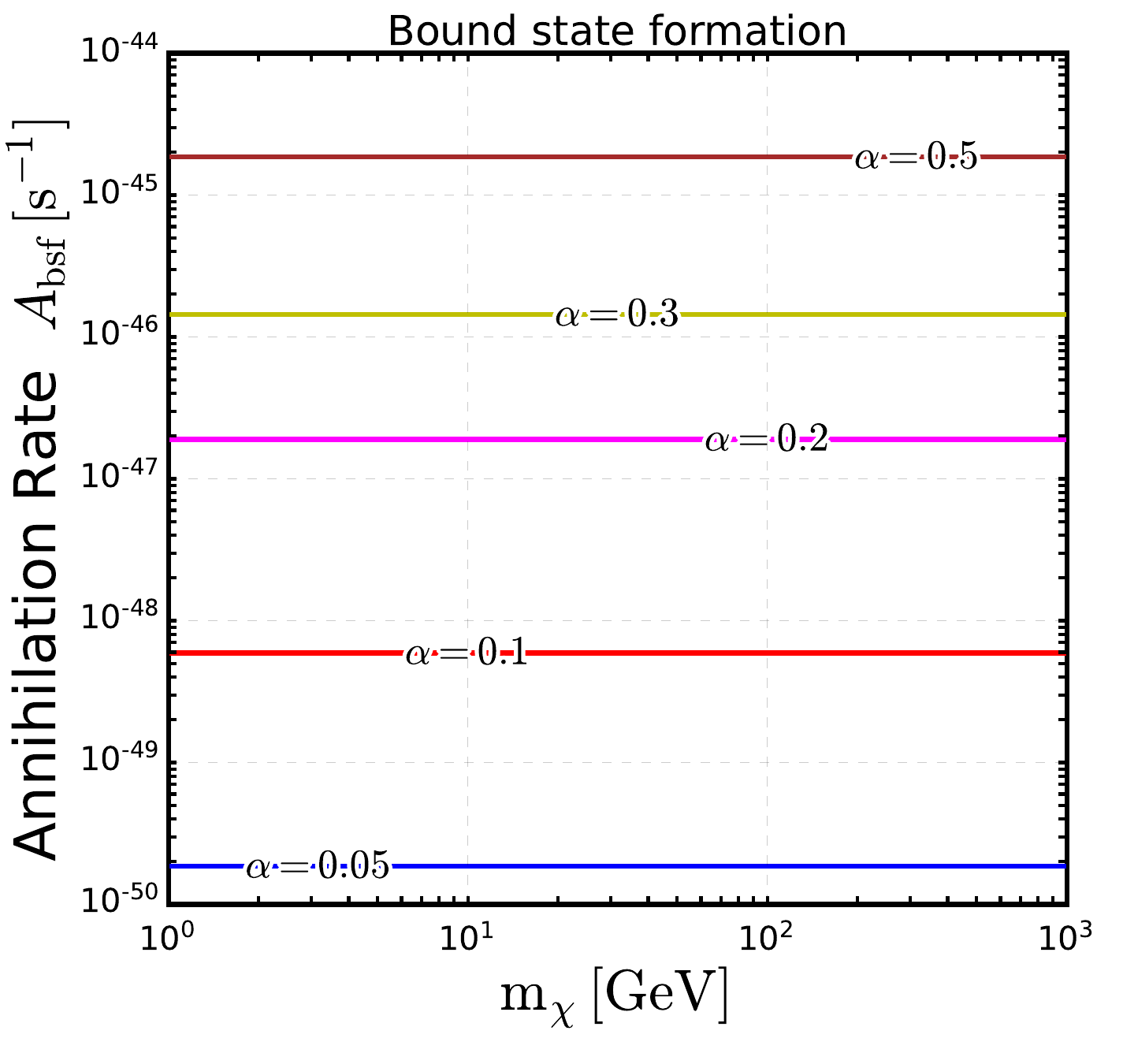}
    \vspace{-1.1cm}
     \caption{\emph{Left}: DM capture rate on nucleons for several mediator masses for a Higgs mixing angle $\sin\theta_\phi = 10^{-10}$. The geometric capture on nucleons $C^g_\star$ and $C_\chi^g$ self-capture rates are shown by the dashed and dashed-dotted black curves. The $C^g_{2\chi}$ self-capture rate line (not shown) is a factor 2 below the $C_\chi^g$ line. \emph{Right}: The DM BSF rate inside the Sun, for several values of $\alpha$.}
     \label{fig:cap_nucl}
%\\
  \vspace{.3cm}
  \includegraphics[width=0.495\textwidth]{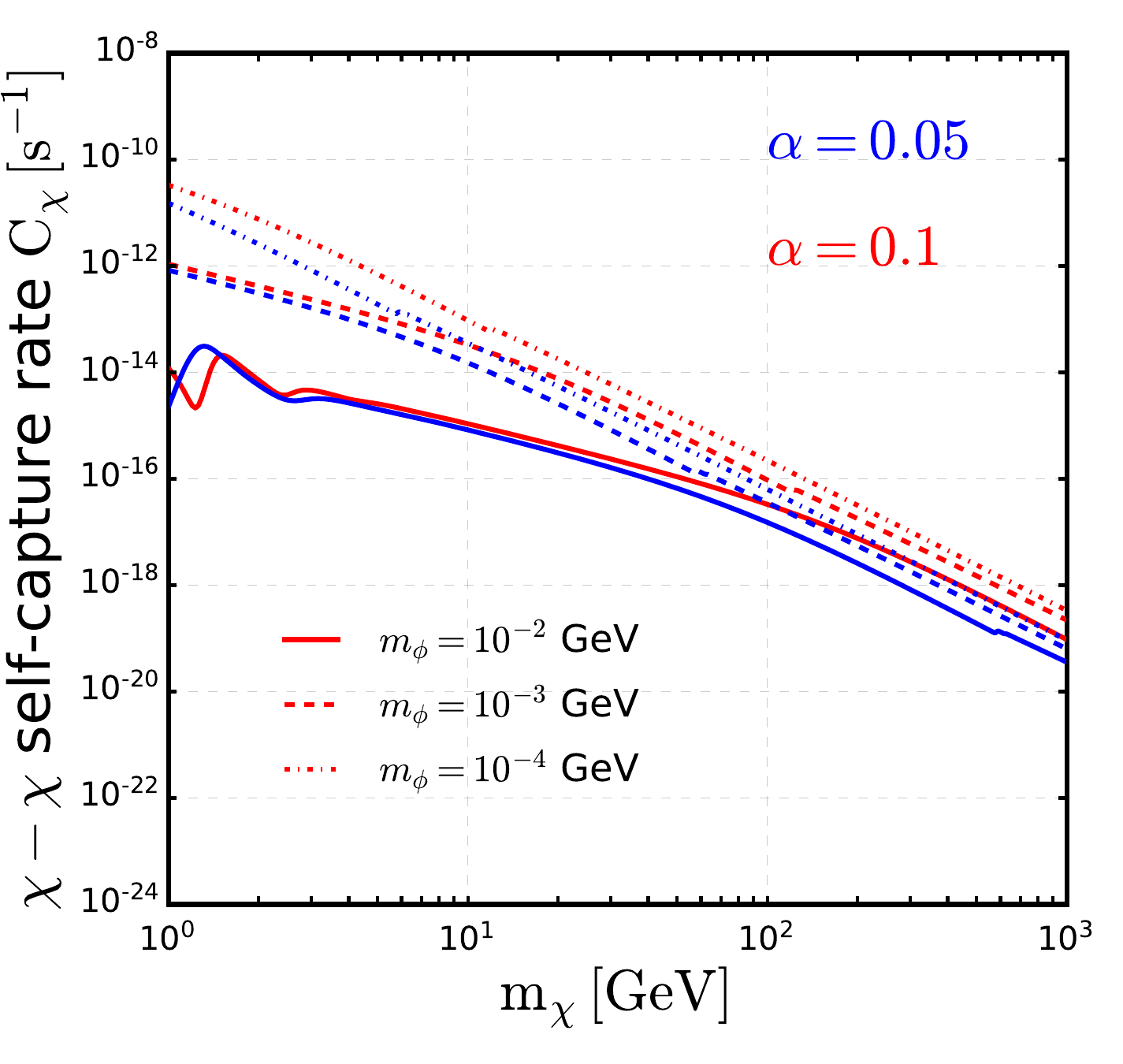}~
 	\includegraphics[width=0.495\textwidth]{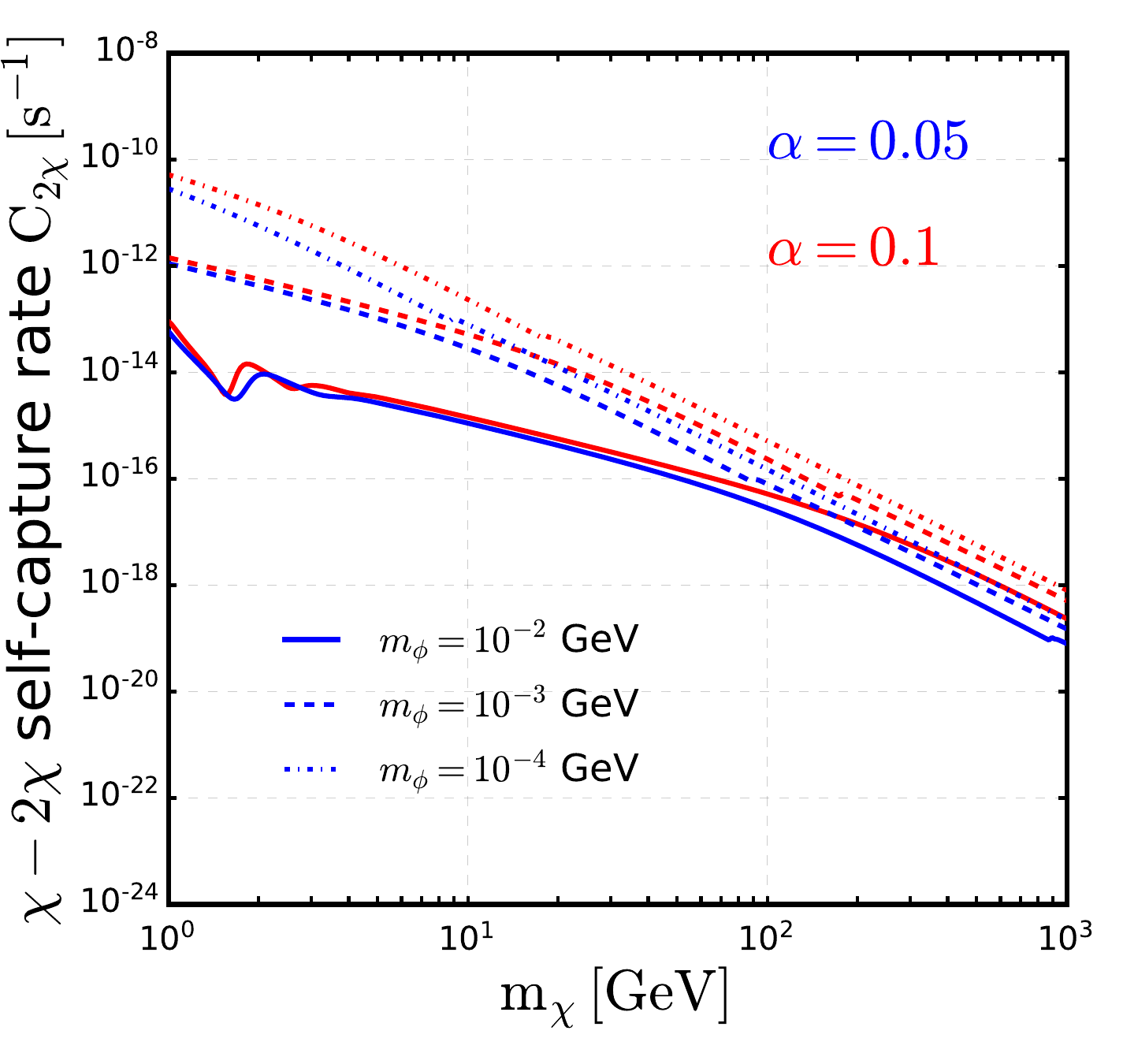}  
  \vspace{-1.1cm}
 \caption{\emph{Left} (\emph{right}): DM self-capture rates due to DM-DM (DM-DMBS) scattering. See text for details. 
 }\label{fig:rates_cs}
\end{figure}

 The DM capture rates on nucleons are shown in Fig.~\ref{fig:cap_nucl}, corresponding to a Higgs mixing angle $\sin \theta_\phi = 10^{-10}$ and $g_s=1$, for several mediator masses (labeled in red).  The geometric capture and self-capture rates, Eq.~\eqref{eq:capturegeom} and \eqref{eq:selfcapgeom}, are shown as black dashed and dotted curves.  For mediator mass below MeV the differential cross section is independent of the mediator mass since $m_N E^{max}_R \sim \mu_{\chi N} v^2_e$, which is  above MeV for $m_\chi \ge m_{\rm evap}$. This leads to a $t$-channel enhancement, which in the $m_\chi \gg m_N$ limit gives a capture cross section scaling as $m^{-1}_\chi$, leading to a capture rate scaling as $m^{-2}_\chi$. While, for mediator masses larger than $\sim 100 $ MeV and $m_\chi \lesssim m_N$, the capture rate in turn scales as $m_\chi$. Note that in Eq.~(\ref{Cstar}) we have neglected the effect of DM-electron scattering, which is subleading due to lower energy loss and tiny electron Yukawa coupling.

%%%%%%%%%%%%%%%%%%%%%%%%%%%%%%%%
\paragraph{\underline{Bound-state formation rate $A_{bsf}$}}

As stated above, we consider DM to be asymmetric and made of Dirac fermions ($\chi$). The interaction between two (identical) DM particles is attractive when the mediator is a scalar ($\phi$), with a classical Yukawa potential $V(r) = - {\alpha \over r} e^{-m_\phi r}\,.$  The binding energy of the bound-state is $E_{\rm bind} \simeq  \frac{m_\chi \alpha^2}{4} - \alpha m_\phi$, with the dark coupling $\alpha\equiv g_s^2/4\pi$. From the merging of two identical fermions, the radiative BSF rate has been estimated in~\cite{Wise:2014jva},  yielding
\beq
\sigma_{\rm BSF} v \simeq  \frac{256 \pi^2 \alpha^5}{5e^4  m_\chi^2 v_{\rm rel} }\,,
\label{BSFrate}
\eeq
in the limit of  $\alpha/v_{\rm rel}\gg 1$ and $m_\phi \to 0 $, where the DM relative velocity is $v_{\rm rel} \simeq 2 v_{\chi, \odot } $. The recoil velocity of the DMBS is $\alpha^2/8$ in this limit, which must be below the escape velocity from the Solar center, resulting in the requirement $\alpha \lesssim 0.18$.\footnote{Due to velocity enhancement, the BSF process in the early Universe could be efficient such that very few free DM particles could exist in the current epoch. Then qualitative changes in some regions of parameter space of interest are expected. However, this depends on the assumed cosmological history, and it is not explored in this work. See e.g.~\cite{Wise:2014jva,Gresham:2017zqi} for 
 related discussions. } Finally, the total annihilation rate that enters the number evolution equation is given by 
\beq \label{eq:Solarbsf}
A_{\rm bsf} = \frac{\int \,n_\chi^2\,\sigma_{\rm BSF} v\, d V }{\left(\int n_\chi d V\right)^2}~.
\eeq
The normalized DM radial distribution is 
\begin{equation}
\label{eq:nDMr}
n_\chi (r, t) = N_{\chi}(t) \, \frac{e^{-m_\chi \phi(r)/T}}{\int_{0}^{R_\odot}  e^{-m_\chi \phi(r)/T} \, 4\pi r^2 \, \dd r} ~,
\end{equation}
which corresponds to an isothermal sphere, with a radial dependence set by the gravitational potential $\phi(r) = \int_0^r G M_\odot(r')/{r'}^2 \, \dd r'$, with $G$ the gravitational constant and $M_\odot(r')$ the mass inside a sphere of  radius $r'$, and $N_{\chi} (t)$ is the total population of DM particles at a given time $t$. 

Once DM particles are thermalized its average velocity is $v_\chi\simeq{} \sqrt{2 T/m_\chi}$. The smallness of this value considerably boosts the BSF in Eq.~(\ref{BSFrate}): $v_{rel}\simeq 4.2 \times 10^{-3} \sqrt{1\,\hbox{GeV}/m_\chi} $. Consequently,  $A_{\rm bsf}$  scales as $\alpha^5 (m_\chi^2 v_{\rm rel} )^{-1} (r^\chi_{\rm th} )^{-3} $,  be approximately independent of the DM mass in practice, as shown in right panel of Fig.~\ref{fig:cap_nucl}.

%%%%%%%%%%%%%%%%%%%%%%%%
\paragraph{\underline{DM self capture rates $C_\chi$ and $C_{2\chi}$}:}

Analogous to DM-nucleon capture rate,  we estimate the self-capture rate per target DM particle in the Sun as follows
\beq \label{eq:cap-xx-self}
C_\chi =    \int_0^{r^\chi_{th}}\dd r 4 \pi r^2 n_\chi(r) \int_0^\infty \dd u_\chi \, \left(\frac{\rho_\chi}{m_\chi}\right) \,u_\chi\, f_\odot(u_\chi)\, w^2 \int_{\cos \theta_{\rm min}}^{\cos \theta_{\rm max}} \frac{\dd \sigma_{\chi-\chi}}{\dd \cos \theta} \,  \dd \cos \theta~.
\eeq
The minimum scattering angle is set by the requirement of a minimum energy that has to be lost in a single scattering event to be captured, i.e.~DM kinetic energy at infinity  must be smaller than $1/2 \,m_\chi u^2_\chi$. The maximum scattering angle for scattering of identical particles is $\pi/2$. Therefore $\cos \theta_{\rm min} =  1- 2 \frac{u^2_\chi}{\omega^2}$  and $ \cos \theta_{\rm max} =  0$~.  Similar to the DM-DM self-capture, the capture rate due to DM-DMBS scattering, per target DMBS in the Sun, is given by
\beq\label{eq:cap-x-2x-self}
C_{2\chi} =  \int_0^{r^{2\chi}_{th}}\dd r 4 \pi r^2 n_{2\chi}(r) \int_0^\infty \dd u_\chi \, \left(\frac{\rho_\chi}{m_\chi}\right) \,u_\chi\, f_\odot(u_\chi)  \, \omega^2 \int_{\cos\theta_{\rm min}}^{\cos\theta_{\rm max}} \frac{\dd \sigma_{2\chi- \chi}}{\dd \cos \theta} \, F_\chi^2\left(\frac{E_r}{Q_\chi}\right) \dd \cos \theta ~,
\eeq
with $\cos \theta_{\rm min} =  1- \frac{9}{8} \frac{u_\chi^2}{\omega^2}$ and $  \cos \theta_{\rm max} =  -1$.
We have used $m_{2\chi} \approx 2 m_\chi$ and set that the coupling to bound-state is $2 \alpha$ (as the vertex is of the scalar type). We assume that the form factor has the form of $\exp(E_r/Q_\chi)$. The typical size of the bound-state is set by the Bohr radius, hence $Q_\chi = m_\chi\alpha$. To a good approximation $F_\chi \rightarrow 1$  for a typical value of $\alpha = 0.1$. The radial number density of DMBS ($n_{2\chi}$) is assumed to be isothermal, analogous to $n_{\chi}$.\\

Bound-state formation suggests that non-perturbative effects are non-negligible for the calculation of the DM self-scattering cross section (see e.g.~\cite{Tulin:2013teo, Colquhoun:2020adl}). This is indeed our case. For the parameter values of interest  $\alpha \in [0.02,0.2]$, $m_\chi > 5$ GeV, $ {\rm MeV} \lesssim m_\phi \ll m_\chi$, and $v \sim v_{esc}\sim 4\times 10^{-3}$, we are never in the Born regime of self-scattering, for which $  m_\chi < m_\phi/\alpha $ or $\alpha \lesssim v $~\cite{Landau:1991wop}. As a result, a perturbative expansion of the scattering cross section is not justified here, see left panel of  Fig.~\ref{fig:regimes} in Appendix~\ref{app:selfscatt}.
However, for these parameters above, it is not necessary to solve the Schr\"odinger equation associated with the non-relativistic DM collisions because the scattering process is typically semi-classical, for which the range of the potential is larger than the DM de Broglie wavelength, that is, $m_\chi v/m_\phi \gtrsim 1$. In this case classical mechanics can be employed to estimate the scattering cross section.\footnote{A precise treatment would require to solve the phase shift for a large number of partial waves, see e.g. ~\cite{Tulin:2013teo, Chen:2015uha, Colquhoun:2020adl}. } 

More importantly, unlike phenomenological studies of DM self-scattering in galaxies and galaxy clusters, where the transfer and viscosity cross sections are relevant, here the differential cross section is needed to calculate the integrated self-capture rates introduced above. This is  because DM capture needs that enough initial kinetic energy is lost in a single scattering. This condition is imposed by integrating the differential cross section from minimum possible scattering angle ($\theta_{\rm min}$, set by required energy loss) to the maximum one where the scattered particle still does not gain enough energy to escape. These limits are explicitly indicated in Eqs.~\eqref{eq:cap-xx-self}
 and~\eqref{eq:cap-x-2x-self}.    To calculate the differential cross sections, we follow  Refs.~\cite{Khrapak:2003kjw, Khrapak:1308514} and  solve the elastic scattering with a Yukawa potential classically. Further discussions and full expressions are presented in Appendix~\ref{app:nonpertscatt}.  

We present the results for DM self capture rates through scattering on  free DM particles (DM bound-states) in the left (right) panel of Fig.~\ref{fig:rates_cs}. The rates scale proportionally as $m^{-2}_\chi$ for $m_\chi \gg m_\phi$. For moderate values of DM masses the scaling is less steep. 
Note that the capture rate on bound-states $C_{2\chi}$ is approximately larger by a factor of two with respect to $C_\chi$, due to larger Yukawa coupling induced by bound-states and larger maximal allowed scattering angle that keep both particles gravitationally captured.

%%%%%%%%%%%%%%%%%%%%%%%%%%%%%%%%%%%%%%%%%%%%%%%%%%
\subsection{Integrating the Boltzmann equations}
\label{sec:2solutions}
 
\begin{figure}[t]%[htb!]
 	\includegraphics[width=0.495\textwidth]{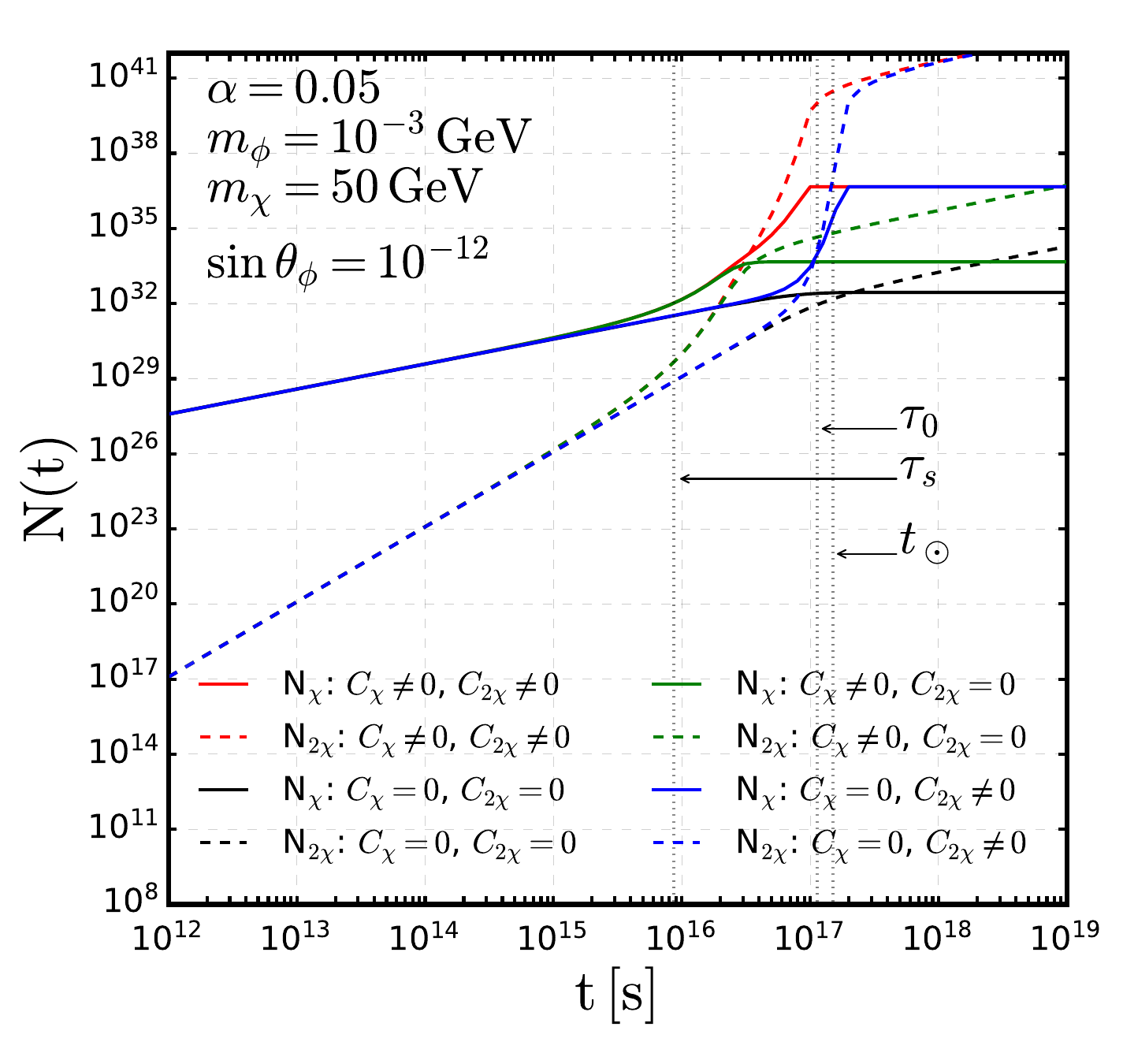}
  \includegraphics[width=0.495\textwidth]{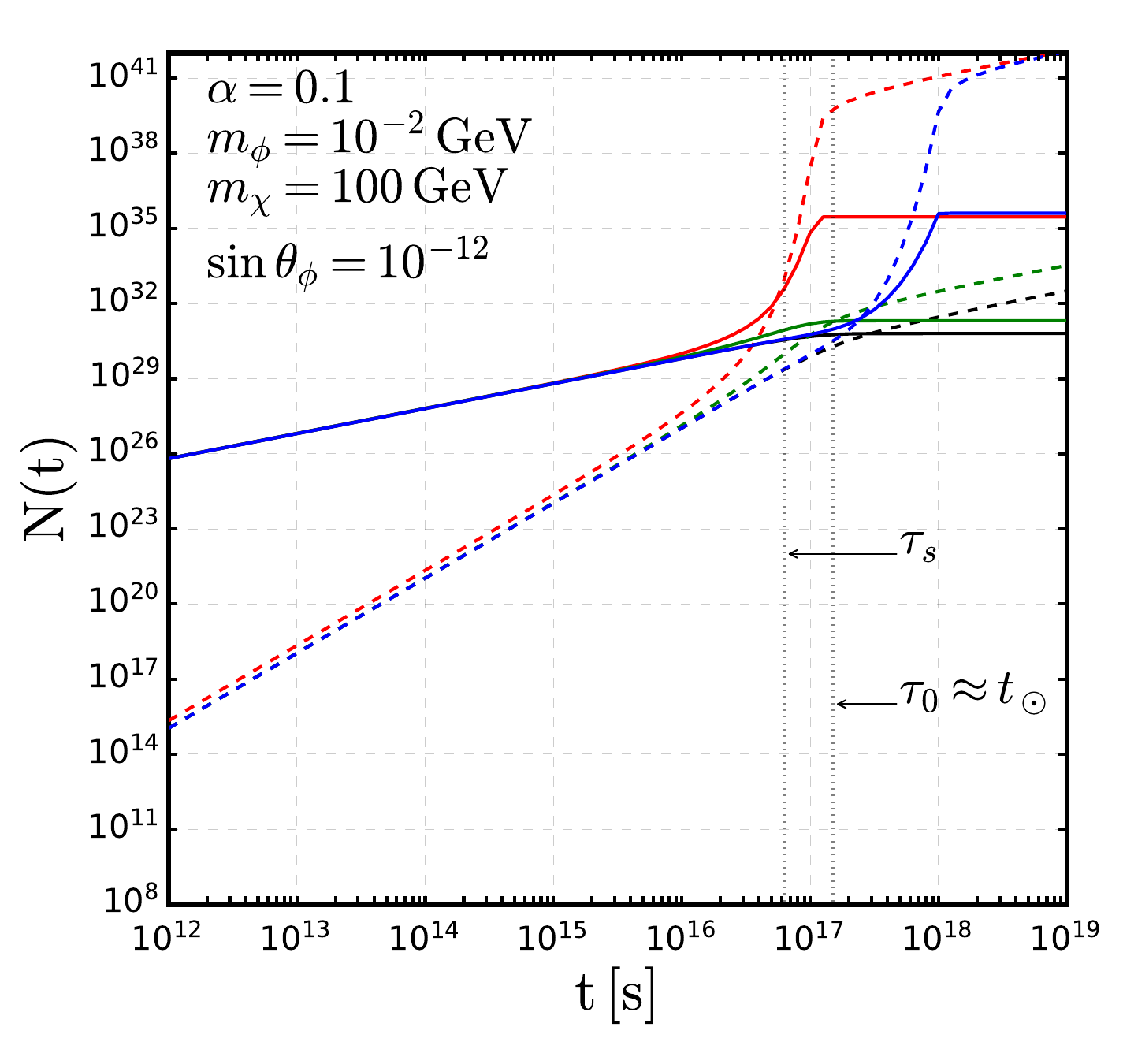}
 \caption{Number evolution of DM and DM bound-states for two parameter sets, as a function of time. Here $\tau_s$ and $\tau_0$ are characteristic times explained in the main text (and exactly defined in Appendix~\ref{App:nx_evolve}), $t_\odot$ is the age of the Sun. The color code of the curves in the left panel are the same as that in the right panel.
 }
	\label{fig:number_evolution}
\end{figure}

The set of coupled Boltzmann equations of Eqs.~\eqref{eq:boltz} and~\eqref{eq:boltz2} has no closed form analytical solution. In  Appendix~\ref{App:nx_evolve} we describe at length how they can be solved approximately. We summarize in this subsection the main outcome of this discussion. To this end, in Fig.~\ref{fig:number_evolution} we present the evolution of both populations for two parameter sets. The evolution of $N_\chi$ ($N_{2\chi}$) from the numerical integration of the full Boltzmann Eqs.~\eqref{eq:boltz} and~\eqref{eq:boltz2}  are given by the solid (dashed) red curves. Curves with other colors correspond to what is obtained switching off both the $C_\chi$ and $C_{2\chi}$ terms (black), or only the $C_\chi$ term (blue) or only the $C_{2\chi}$ term (green).

At early times the number of free particles $N_\chi$ grows as $C_\star \, t$ (i.e.~red and black solid lines coincide). As $N_\chi$ increases, the BSF starts to occur efficiently and the corresponding term (quadratic in $N_\chi$) quickly catches up with the constant term associated with capture on nucleons, so that in absence of the $C_{2\chi}$ term, a  quasi-static equilibrium between both terms is reached (see black and green solid curves in Fig.~\ref{fig:number_evolution}). Solving the Boltzmann equation for $N_\chi$ one gets that the associated time scale is $\tau_0  \equiv (C_\star A_{\rm bsf})^{-1/2}$ if one drops the $C_\chi $ term --which has a subleading effect (black solid curve)-- or $\tau_s \equiv (C_\star A_{\rm bsf} +C^2_\chi/4)^{-1/2}$ if one includes it (green solid curve). For more details,  see the end of this section and Appendix~\ref{App:nx_evolve}.

Another consequence of efficient BSF is that the number of bound-states $N_{2\chi}$ quickly catches up to the number of DM particles $N_\chi$, and never stops growing thereafter, as it is not counterbalanced by any other term in Eq.~\eqref{eq:boltz2}.  From this point, the term associated with  the capture on bound-states, $C_{2\chi} N_{2\chi}$, takes over and leads to an exponential growth of $N_{\chi}$ (red and blue solid curves  in Fig.~\ref{fig:number_evolution}) and of $N_{2\chi}$ (dashed curves with same colors). The exponential growth become significant at $t \gtrsim \tau_s  +  C_{2\chi}^{-1}$, or at $\tau_0+  C_{2\chi}^{-1}$, depending on whether one includes the $C_\chi$ term. This  growth lasts until the capture rate saturates the geometric rate within the bound-state thermal sphere, i.e.~when the  term $C_{2\chi} N_{2\chi}$ reaches $C^g_{2\chi}$. This happens when the mean free path of DM particles becomes much smaller than the bound-state thermal radius.   At this moment, we can safely neglect the  $C_{\chi}N_\chi$ term  as $N_\chi \ll N_{2\chi}$,  thus the Boltzmann equation takes the form, $d N_\chi/dt \simeq C_\star+C^g_{2\chi}  -A_{\rm bsf} N_\chi^2$.
 This means that, quickly after the geometric capture rate is saturated,  the BSF term  compensates the constant capture rate from both nucleon and bound-state terms. Setting $d N_\chi/dt  \simeq 0 $ gives the maximum final value 
\beq\label{eq:sol_eqbn}
N_{\chi,{\rm eq}} \simeq \left( \frac{C_\star +  C_{2\chi}^g   }{A_{\rm bsf}}\right)^{1/2}~ .
\eeq
We denote as $\tau_g$ the time when $N_\chi$ freezes in such a way. An analytic approximation is given in Appendix~\ref{App:nx_evolve}.

Note that,  contrary to the $C_{2\chi}N_{2\chi} $ term --which saturates the geometric rate $C_{2\chi}^g$--  the average DM density within the  thermal radius $r^\chi_{\rm th}$ is smaller than the DMBS density within $r^{2\chi}_{\rm th}$, so the self-capture term $C_{\chi} N_{\chi}$ does not reach $C^g_{\chi}$ at  $t=\tau_g$, and will never reach it after, as $N_{\chi}$ stops increasing. As for $N_{2\chi}$,  it keeps increasing in time forever. For $t>\tau_g$, it increases linearly in time and the rate of BSF is half the capture rate,
\beq
\label{gammaflux}
\Gamma(t) = \frac{d N_{2\chi}}{dt} = \frac{1}{2} A_{\rm bsf} N^2_{\chi,{\rm eq}} \simeq   \frac{C_\star+C^g_{2\chi}}{2}  \,,
\eeq
inducing a flux of mediators given by the same rate.

The reason why the $C_\chi$ term has a subleading effect with respect to the effect of the $C_{2\chi}$ term stems from the fact that these two terms are very different in nature. If we switch off the $C_{2\chi}$ term (green curves), the $C_\chi$ term also leads to an exponential grow, starting slightly before $t=\tau_s$, but it is much less important than the exponential growth from the $C_{2\chi}$ term. This is analogous to what happens for the symmetric DM case with BSF playing the role of annihilation if $C_{2\chi}=0$.\footnote{In the limit of $C_{2\chi} \rightarrow 0$ and $A_{\rm bsf}=0$ (and with an additional DM annihilation term), the solution has an analytical form, first presented in~\cite{Zentner:2009is}. For more recent works, see e.g. \cite{Catena:2016ckl, Gaidau:2018yws,Gaidau:2021vyr}. }
In Eq.~(\ref{eq:boltz2}) the $C_\chi$ term (linear in $N_\chi$) is quickly counter-balanced by formation of bound-states from the $A_{\rm bsf}$ term (quadratic in $N_\chi$): after a short period of exponential growth, $N_\chi\sim C_\star (e^{C_\chi t}-1)/C_\chi$, the green curve ($C_{2\chi}=0$),  goes to a plateau from an equilibration of capture and BSF. 
In contrast, if $C_{2\chi}\neq 0$, the capture rate on bound-state grows as $N_{2\chi}$. Thus, due to capture on bound-state term $C_{2\chi}$, if it were not for the geometric rate upper floor, both $N_\chi$ and $N_{2\chi}$ would grow forever. Moreover,  the argument of the exponential  from the $C_{2\chi}$ term is larger than the one from the $C_\chi$ term, because $C_{2\chi}$ is about a factor of two times $C_\chi$, see Fig.~\ref{fig:rates_cs}.

In presence of both $C_\chi$ and $C_{2\chi}$ terms, the $C_\chi$ term has only a moderate impact on the evolution of $N_\chi$ and $N_{2\chi}$. As a comparison of the full evolution (red curves  in Fig.~\ref{fig:number_evolution}) and the  $C_\chi=0$ evolution (blue curves) shows, including $C_\chi$ reduces the timescale when the exponential growth starts, from $t\sim \tau_0$ to $t \sim \tau_s$. Thus, due to the $C_\chi$ term, the exponential effect of the $C_{2\chi}$ term starts somewhat earlier, and the geometric rate within the  bound-state thermal sphere is also reached somewhat earlier. The values of $N_\chi$ and $N_{2\chi}$ are insensitive to the $C_\chi$ term at $t\gg \tau_s$.

%%%%%%%%%%%%%%%%%%%%%%%%%%%%%%%%%%%%%%%%%%%%%%%%%%%
%%%%%%%%%%%%%%%%%%%%%%%%%%%%%%%%%%%%%%%%%%%%%%%%%%%

\section{Neutrino flux and Terrestrial  detection}
\label{sec:neutrinoflux}

As explained above, thanks to the capture on DMBS, the capture rate will saturate the geometric rate within the DMBS thermal sphere, leading to an equilibrium between the capture and the BSF processes. The time $\tau_g$, at which this happens, can be easily shorter than the age of the Sun, as solved in Eqs.~(\ref{tau1}) and (\ref{tau1versustauSun}). At this point, the flux of mediators emitted from BSF becomes constant and is approximately given by Eq.~(\ref{gammaflux}) above.

\begin{figure}[t]%[htb!]
	\includegraphics[width=0.495\textwidth]{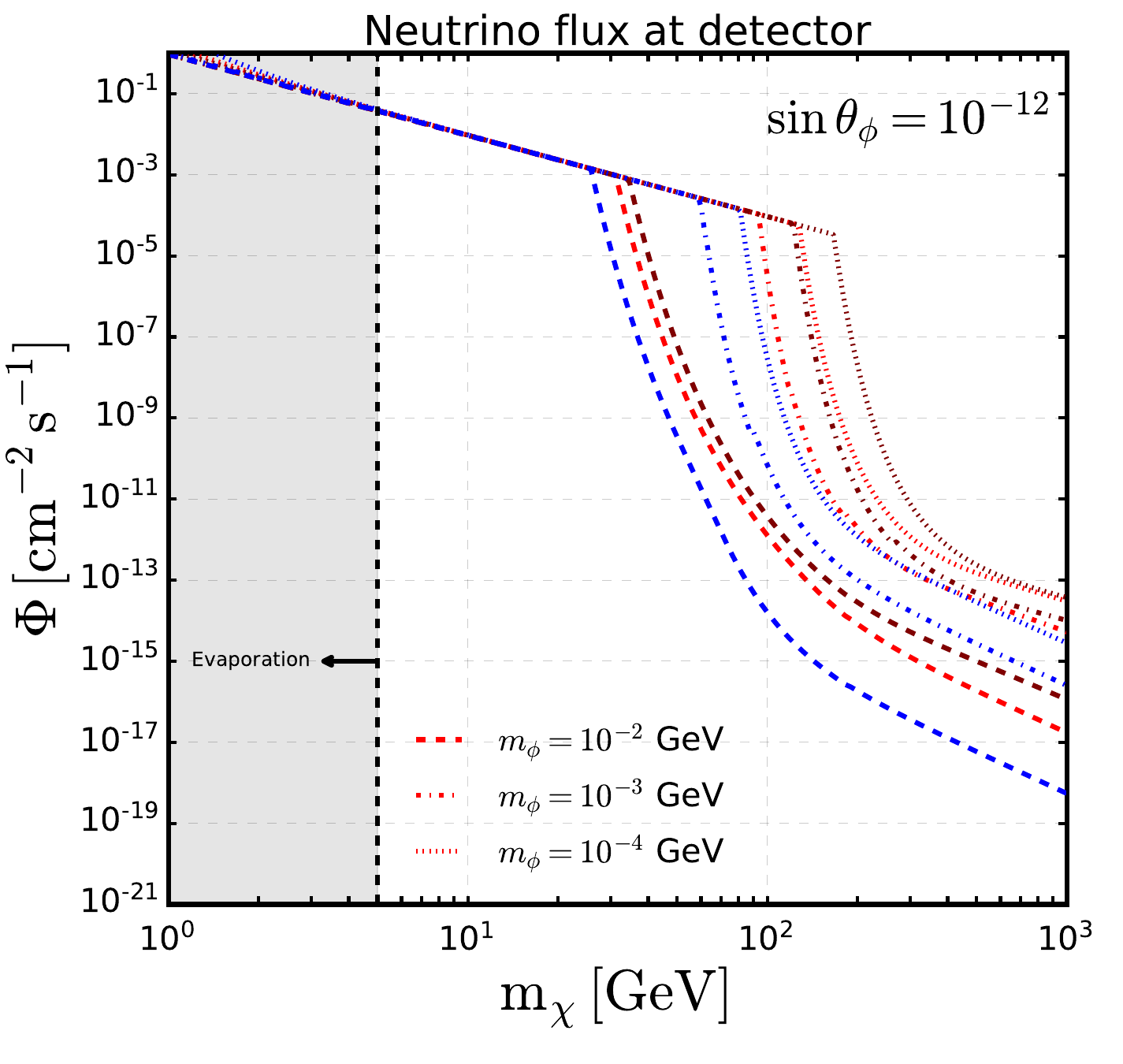}~~    \includegraphics[width=0.495\textwidth]{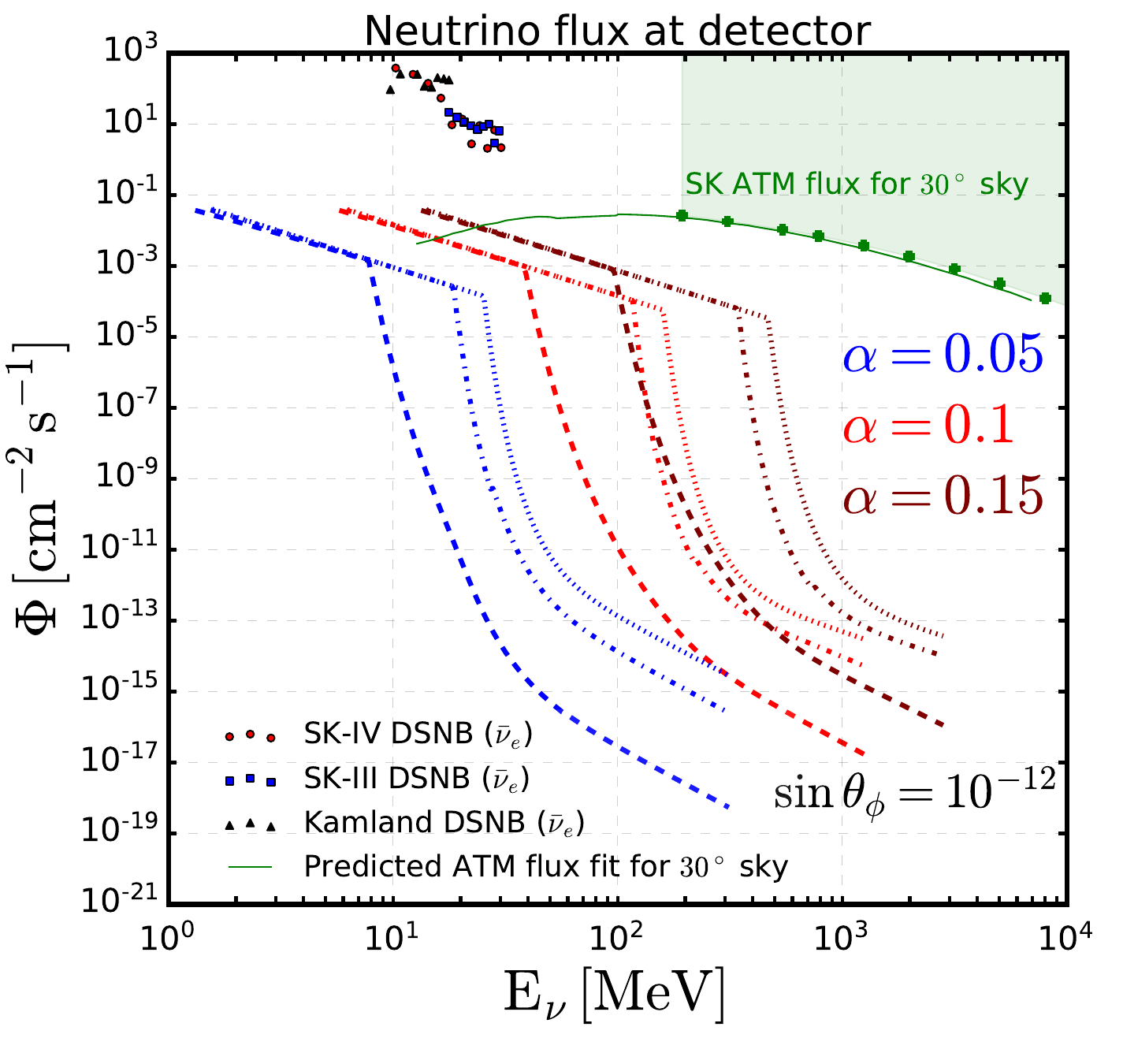}
 %%%   
 \caption{\emph{Left}: Flux of light mediators emitted by the BSF process for various parameter choices. \emph{Right}: Neutrino flux as a function of neutrino energy, coming from the Sun for a detector placed on the surface of the Earth. Shown in red, blue and black scattered points are the current limits on diffuse supernova neutrinos adapted from~\cite{Super-Kamiokande:2021jaq}. Atmospheric low energy neutrino measurements from Super-K are shown by the green dots~\cite{Super-Kamiokande:2015qek}. Thin green line denotes the predicted atmospheric neutrino flux for $30^\circ$ sky~\cite{Suliga:2023pve}.  
 }
	\label{fig:flux_cs}
\end{figure}

For concreteness (and other reasons explained below) we will assume in the following that the mediator decays dominantly into a pair of neutrinos.
Under the assumption that the mediator mass is much smaller than the binding energy, $E_{\rm bind}$, the differential flux at a terrestrial detector  is
\beq
{d^2\Phi_\nu \over dE_\nu d\Omega}  = {A_{\rm bsf} N_\chi^2(t_\odot)  \over 2 \Delta\Omega} {dN_\nu \over dE_\nu}= {A_{\rm bsf} N_\chi^2(t_\odot)  \over 8 \pi d_\odot^2 \left(1- \cos\vartheta \right)} {dN_\nu \over dE_\nu}\,,
\eeq
where $d_\odot$ is the Earth-Sun distance (AU), with $\vartheta$ being the angular sensitivity of detector (as the apparent angular diameter of the Sun $\vartheta_\odot \approx 0.5^\circ$). 
Since the light mediator is boosted, the neutrino energy spectrum this decay leads to is not monochromatic but has a characteristic box shape~\cite{Ibarra:2012dw} 
\begin{align}
{dN_\nu \over  dE_\nu}  \simeq   {2 \over  E_{\rm bind}} \text{~~~for~}E_{-}\le E_\nu \le E_{+}\,,
&&
\text{with}
&&
E_{\pm}=\frac{E_{\rm bind}}{2}\Big(1\pm\sqrt{1-\frac{4m^2_\phi}{E^2_{bind}}}\Big)\,.
\end{align}
In particular, $E_- \simeq 0 $  and $E_+ \simeq  E_{\rm bind}$  if  $m_\phi \ll E_{\rm bind}$.
The differential flux of neutrinos at the detector (ignoring oscillations) is
\begin{eqnarray}\label{eq:spectrum}
    \frac{d \Phi_\nu}{d E_\nu} &= & 
    \frac{1}{4 \pi d^2_\odot} \Gamma(m_\chi,m_\phi,\alpha_\chi, s_\theta) \frac{2}{E_{\rm bind}}\Theta(E_+ - E_\nu) \Theta(E_\nu - E_-)~.
\end{eqnarray}
The resulting neutrino fluxes are shown in Fig.~\ref{fig:flux_cs} for a fixed Higgs mixing angle $\sin \theta_\phi = 10^{-12}$, for $\alpha=0.05,\,0.1$ and $0.15$ (in blue, red and  maroon colors), respectively.  For comparison, the current limits (90\% C.L.) on diffuse supernova $\nu_e$ neutrino background (DSNB) flux from Super-K runs III, IV and KamLAND are shown in blue, red and black points~\cite{Super-Kamiokande:2015qek,Super-Kamiokande:2021jaq, KamLAND:2021gvi}. Also shown are predicted atmospheric fluxes of $\nu_e$ down to 10 MeV in thin green line, assuming $30^\circ$ angular resolution in the sky, adapted from~\cite{Battistoni:2005pd,Zhou:2023mou,Suliga:2023pve}. Current measurements of atmospheric neutrino fluxes are shown in green points~\cite{Honda:2015fha}. As this figure suggests,  the neutrino flux induced by BSF could be observable. When the geometric limit for DMBS sphere is reached, the emitted flux of mediators is proportional to $C^g_{2\chi} \propto m^{-2}_\chi$. This scaling is seen in Fig.~\ref{fig:flux_cs}. If the geometric limit is not reached, heavier mediators would lead to smaller self-capture rates, as shown by the falling tails of the neutrino flux. 

Finally, note that our results can be easily generalized, e.g. if the mediator decays electromagnetically. Nevertheless, the experimental constraints, as studied below, also become more stringent for  electromagnetic decays, further narrowing down allowed regions of parameter space. 

%%%%%%%%%%%%%%%%%%%%%%%%%%%%%%%%%%%%%%%%%%%%%%%%%%%%%%%%
%%%%%%%%%%%%%%%%%%%%%%%%%%%%%%%%%%%%%%%%%%%%%%%%%%%%%%%%
\section{Results and constraints for an explicit model}
\label{sec:bounds}

%%%%%%%%%%%%%%%%%%%%%%%%%%%%%%%%%
\subsection{Model}

As mentioned above,   we  consider DM in the form of a (vector-like)  Dirac fermion self-interacting through the exchange of a light scalar.
Concretely the Lagrangian is (see e.g.~\cite{Krnjaic:2015mbs,Winkler:2018qyg,Matsumoto:2018acr})
\begin{eqnarray}
{\cal L} &\supset& {\cal L}_{\rm SM} +  {\cal L}_{N_R}+  \bar{\chi}\left(i\slashed{\partial}  -m_\chi \right)\chi + \frac{1}{2} \left(\partial \phi \right)^2 - g_s \phi \bar{\chi} \chi  -V(\phi,H)\,.
\end{eqnarray}
 Here we do not specify the scalar potential $V(\phi, H)$, and simply assume that it induces a $\phi$-$H$ mixing angle $\theta_\phi$ (which can be achieved in various ways), so that, 
upon rotation to  the mass basis, the following interactions are obtained:
\begin{eqnarray}
{\cal L}_{\rm int}& = & g_s \cos\theta_\phi \phi^\prime  \bar{\chi} \chi + g_s \sin\theta_\phi h^\prime  \bar{\chi} \chi  +\left(\cos\theta_\phi h^\prime + \sin\theta_\phi \phi^\prime\right)\frac{m_f}{v_H}\bar{f}_L f_R \, .
\end{eqnarray}
If we assume that neutrino masses are generated through the usual type-I seesaw mechanism, nothing prevents the right-handed neutrinos to couple in pairs to the light mediator $\phi$. Thus one has the extra interactions
\begin{eqnarray}
  -{\cal L}_{N_R} & =&  Y_\nu \bar{N} L\cdot H + \frac{m_N}{2} \overline{N^c} N +  Y_\phi\phi \overline{N^c} N   + {\rm h.c.} ~.
  \label{LagrN}
\end{eqnarray}
 In the physical $\nu', N'$ and $\phi', h'$ mass eigenstate basis, this leads to the following relevant Yukawa interactions for the light scalar eigenstate $\phi'$, up to second order in the neutrino mixing angle,
\begin{eqnarray}
  -{\cal L}_{\rm N_R}& \supset & 
Y_\phi \phi' \cos \theta_\phi 
(\overline{N'^c}N' \cos^2\theta_\nu
  +\overline{N'^c} \nu'^c  \sin \theta_\nu
    +\overline{\nu'} N'  \sin \theta_\nu
    +\overline{\nu'}\nu'^c \sin^2\theta_\nu)+h.c.\nonumber\\
&-&Y_\nu\phi' \sin \theta_\phi 
(\bar{N'}\nu' \cos^2\theta_\nu 
+\overline{\nu'^c} \nu'  \sin \theta_\nu
-\bar{N'} N'^c \sin \theta_\nu
-\overline{\nu'^c}N'^c \sin^2 \theta_\nu)+ h.c.\,.
   \label{LagrNdiag}
\end{eqnarray}
Thus, both the $Y_\phi$ and $Y_\nu$ interactions induce a decay of the light mediator into a pair of light neutrinos. In practice the decay induced by $Y_\phi$ will be dominant because various constraints require $Y_\nu \sin \theta_\phi \ll Y_\phi \sin \theta_\nu$. 
In the following, for simplicity, we will consider only one left-handed neutrino and one right-handed  Majorana neutrino, with the neutrino mixing angle $\sin \theta_\nu \approx Y_\nu v_h/\sqrt{2}M_N$. Likewise for the associated Yukawa coupling, the typical seesaw gives
\begin{equation}
\sin^2\theta_\nu \sim \frac{m_\nu}{m_N}=10^{-13}\cdot\frac{m_\nu}{0.1 \hbox{eV}}\cdot \frac{1\, \hbox{TeV}}{m_N}
\label{sinalpha}
\end{equation}
with $Y_\nu\sim \sqrt{2 m_\nu m_N/v_h^2}$.

If $m_N>m_\phi$ the light mediator cannot decay into $NN$ or $N\nu$ and the $\nu\nu$ is the only possible channel induced by the seesaw interactions. The decay width is 
\begin{equation}
\label{eq:widthnu}
\Gamma(\phi\rightarrow \nu \nu)=       \frac{1}{16\pi} |Y_ \phi|^2 \, m_\phi \,\sin^4\theta_\nu    \,.
\end{equation}
In addition to this neutrino-mixing channel, $\phi$ can also decay,  through the $\phi$-$H$ scalar mixing, into pairs of charged leptons or quarks (or SM bosons for large $\phi$ mass).  As will be discussed below, this needs to be suppressed.

%%%%%%%%%%%%%%%%%%%%%%%%%%%%%%%%%%%%%%
\subsection{Constraints and results}

An observable neutrino flux requires sufficiently large capture and BSF rates, eventually leading to an equilibrium between both processes. This scenario must also fulfill additional phenomenological constraints:

\begin{figure}[t]
	\includegraphics[width=0.495\textwidth]{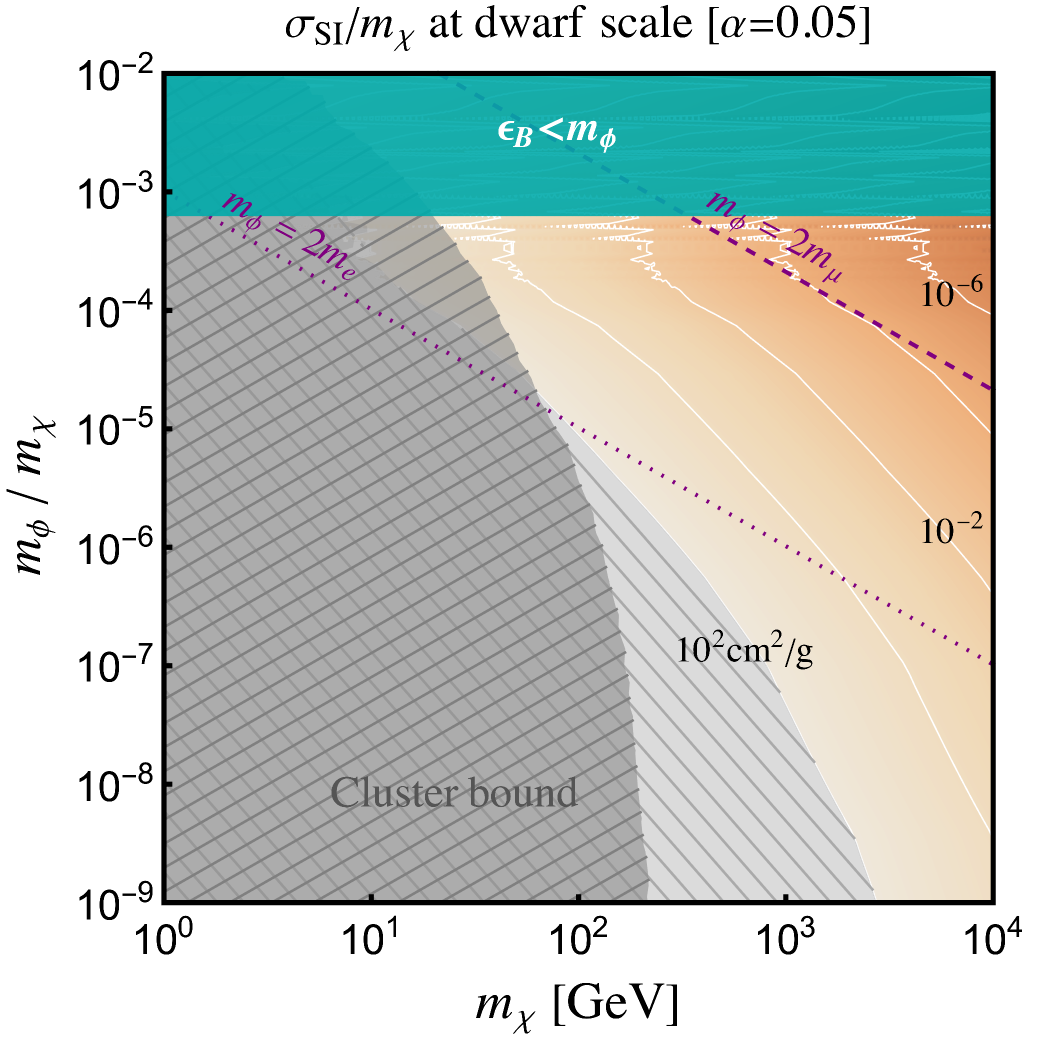}  
 \includegraphics[width=0.495\textwidth]{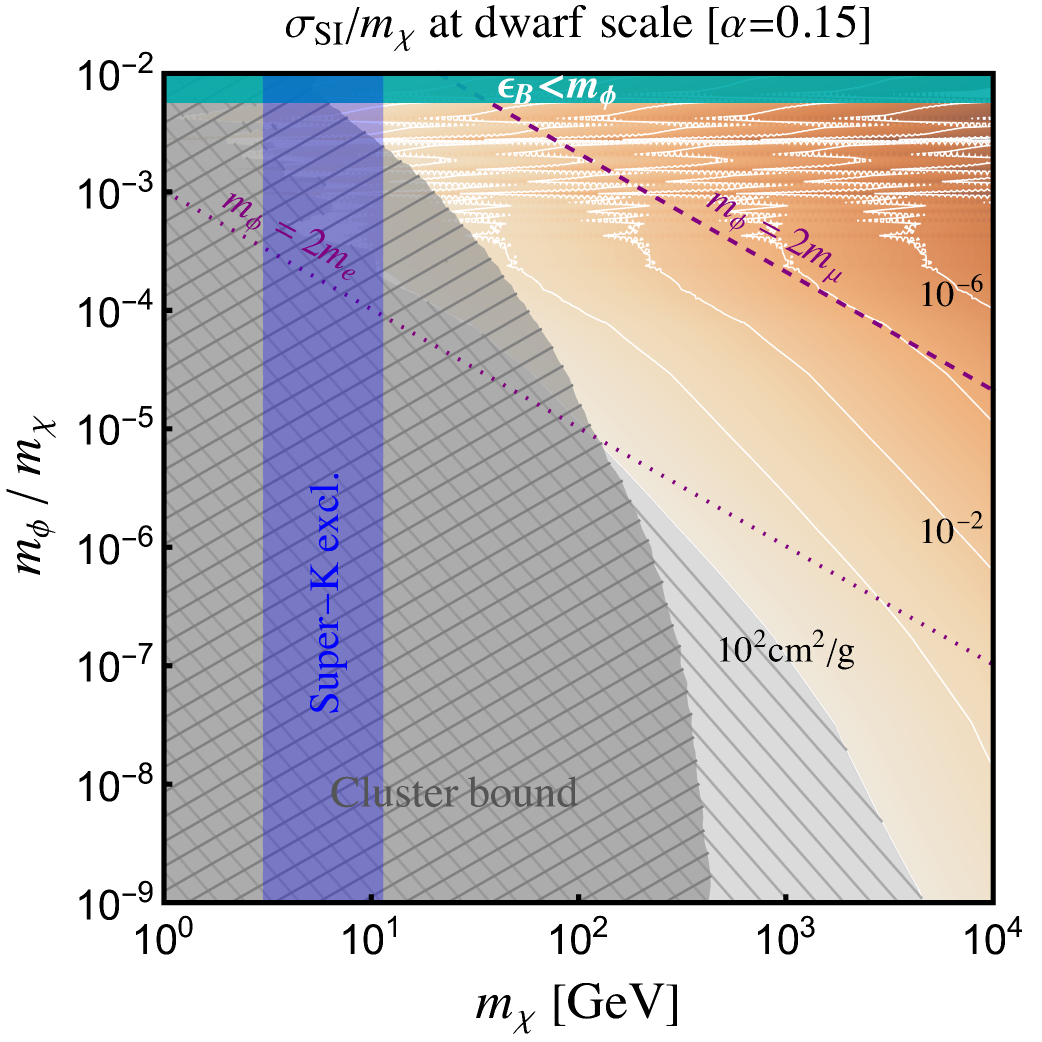}  
	\caption{Constraints for two example values of the dark coupling $\alpha$.
 Contour lines give the values of $ \sigma_{\rm SI} /m_\chi$ in dwarf-sized halos.  Black (gray) hatched regions are excluded because  $ \sigma_{\rm SI} /m_\chi  \geq 0.5$\,cm$^2$/g at cluster scales (or 100\,cm$^2$/g at dwarfs scales, see contour lines).   The resonant regime (with peaks) will become smaller in the Milky Way halo, where the average DM velocity is much larger than that at dwarfs scale.
 }
	\label{fig:SIDM}
\end{figure}

\underline{\it Self-interactions}:  
On the one hand, the corresponding self-scattering cross section is bounded from above by the observation of galaxy cluster collisions, e.g.,~\cite{Harvey:2015hha, Bondarenko:2017rfu, Harvey:2018uwf, Sagunski:2020spe, DES:2023bzs}
\begin{equation}\label{eq:Bullet}
	{\sigma_{\rm SI} \over m_\chi}\,\lesssim \, 0.5\,\text{cm}^2/\text{g}~,
\end{equation}
for DM velocities around $v \sim  1000\,$km/s. On the other hand, the non-observation of gravothermal collapse in dwarf-sized halos sets an upper bound of about $100\,\text{cm}^2/\text{g}$ at  small scales, for which we take $v\sim   25\,$km/s~\cite{Tulin:2017ara, Adhikari:2022sbh}.  Concretely,  we derive the bounds taking $\sigma_{\rm SI}$ equal to the viscosity cross section of  DM particles. Adopting the modified transfer cross section barely changes the results, see Appendix~\ref{app:selfscatt} for detailed discussions of these effective cross sections. 

\begin{figure}[t]
 \includegraphics[width=0.50\textwidth]{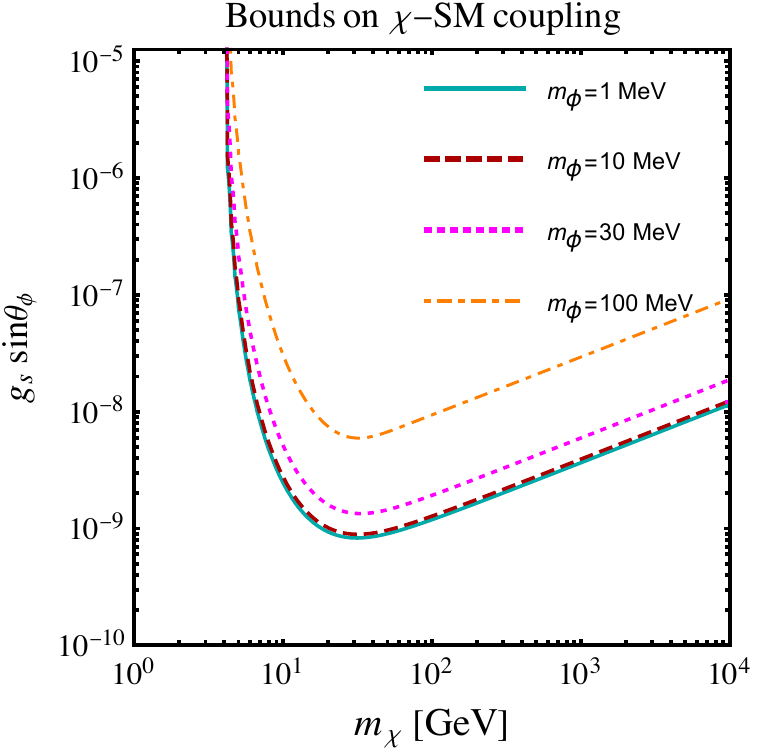}
 \includegraphics[width=0.49\textwidth]{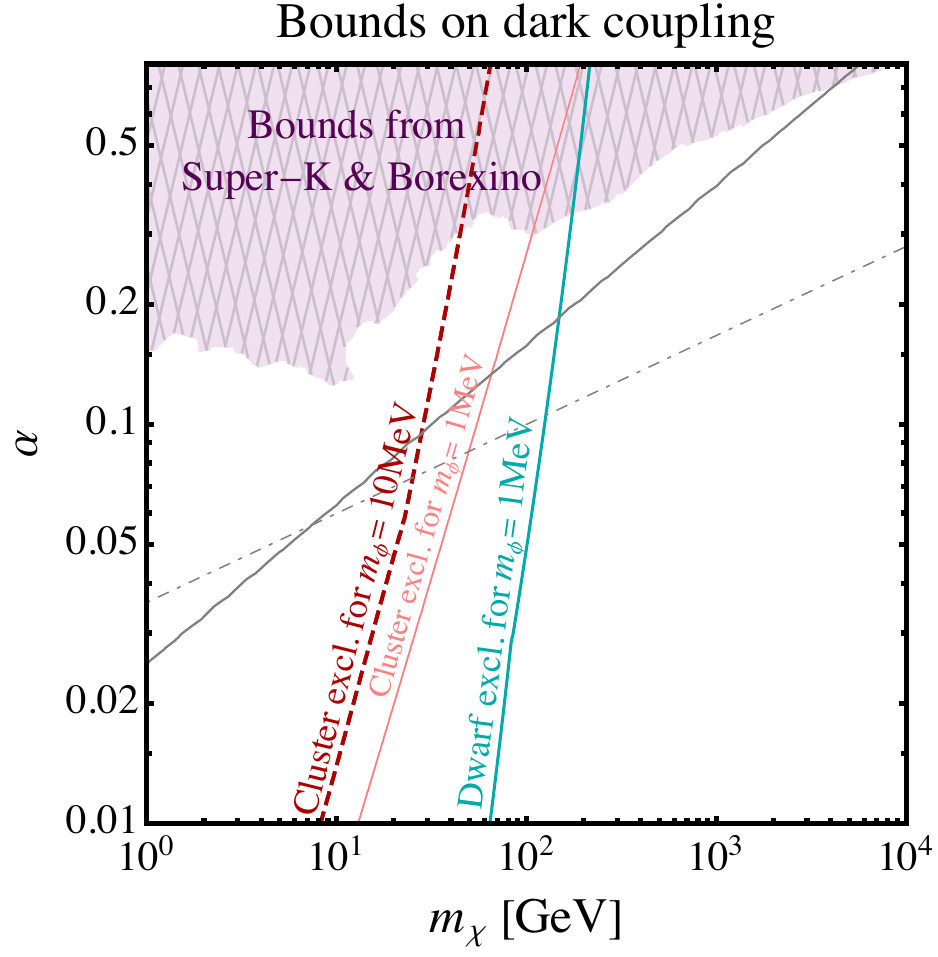}
	\caption{\emph{Left}: Upper bounds on the DM-SM interaction strength from XENON1T~\cite{XENON:2018voc}, depending on $m_\phi$.
 \emph{Right}:  Upper bounds on DM coupling $\alpha$ from indirect search for neutrino flux (purple shaded area) and DM self-scattering at dwarfs scales. 
 The bounds from the Bullet cluster, Eq.~(\ref{eq:Bullet}), are given for $m_\phi = 1\,$MeV and $10\,$MeV. The conservative dwarf scale upper bound, ${\sigma_{\rm SI} / m_\chi}\,\lesssim \, 100\,\text{cm}^2/\text{g}$ is given for $m_\phi=1$~MeV. 
 This case is disfavored by BBN observables, given its sizeable coupling to neutrinos.
 For $m_\phi = 100\,$MeV, the dwarf galaxy scale bound is weaker and basically irrelevant for BSF. For $\alpha$ above the gray solid line,  one expects a galactic flux from BSF in the galactic halo larger than the flux from BSF in the Sun. Efficient BSF in the early Universe is possible for parameters above the dot-dashed gray line~\cite{Gresham:2017zqi}.    }
	\label{fig:indirectB}
\end{figure}

Fig.~\ref{fig:SIDM} shows the resulting values for this cross section as a function of $m_\chi$ and $m_\phi/m_\chi$ for two  values of $\alpha$. The cluster bound excludes  $\chi$ masses below $\sim 10$-$1000$~GeV depending on $m_\phi/m_\chi$ and $\alpha$. This figure also shows the region where  $m_\phi$ is larger than the binding energy where BSF is not possible.  Values of the cross section that could address the small scale anomalies are allowed, i.e.~$\sigma_{\rm SI}/m_\chi\in [1, \,100]  \,\text{cm}^2/\text{g}$ at typical DM velocities of the order of  25\,km/s in dwarf galaxies~\cite{Vogelsberger:2012ku, Zavala:2012us, Elbert:2014bma, Ren:2018jpt}. In the $m_\chi$-$\alpha$ plane, Fig.~\ref{fig:indirectB} (right) shows the corresponding upper bound on $\alpha$.  

%%%%%%%%%%%%%%%%%%%%%%%%%%%%%%%%%%%%%%%%%%%%%%%%%%%%
%%%%%%%%%%%%%%%%%%%%%%%%%%%%%%%%%%%%%%%%%%%%%%%%%%%%

\underline{\it BBN}: In order to observe the neutrino flux in the detector, the  in-flight lifetime of the mediator must be smaller than eight minutes. This constraint is typically less stringent than that resulting from  upper bounds on extra radiation after neutrino decoupling, which requires its lifetime to  be shorter than one second. 
Using the typical seesaw expectation of Eq.~(\ref{sinalpha}), one gets
\begin{equation}
|Y_\phi|^2\sim   0.03 \,\, \frac{100\,\hbox{MeV}}{m_\phi}\frac{1\,\hbox{s}}{\tau_\phi} \,\Big(\frac{0.1 \hbox{eV}}{m_\nu} \frac{m_N}{1 \hbox{GeV}} \Big)^2\,.
\label{nunuchannel}
\end{equation}
Thus a fast enough decay is obtained provided that $Y_\phi$ is not too small and $m_N$ is not too large. BBN and perturbative couplings, $Y_\phi^2\lesssim 4 \pi$,  require $m_N\lesssim  20\,  \hbox{GeV}\cdot (m_\nu/0.1\,\hbox{eV})\cdot (m_\phi/100\,\hbox{MeV})^{1/2}$. Thus low scale seesaw is favored along this scenario. This BBN bound can be relaxed if the $\phi$ number density gets suppressed  before neutrino decoupling.

\underline{\it Direct detection}: As already mentioned above, due to evaporation, DM masses below a few GeV are not relevant for our purpose. Thus, non-observation of spin-independent nucleon recoil signals provides the best direct-detection bounds. For purely scalar-mediated interactions, the differential elastic scattering cross section with nuclei has been given by Eq.~\eqref{Cstar}. If $m_{\phi}\lesssim \sqrt{2 m_N E_R}\lesssim \unit[1]{MeV}$ for typical keV-scale recoil energies, direct detection rates, albeit in the Born regime, are boosted due to $t$-channel exchange of the light mediator, leading to more stringent bounds than for heavier mediators. The corresponding measurements by the Xenon1T experiment~\cite{XENON:2018voc, GAMBIT:2018eea} set an upper bound on $g_s \sin \theta_\phi$, which we show in Fig.~\ref{fig:indirectB} (left) for different masses of mediator and DM. This obviously translates into an upper bound on the capture rate on nucleons, as both processes involve the same DM-nucleon cross section.  Recent LZ data can improve the limit on $g_s \sin \theta_\phi$ by a factor of $2$\,--\,$3$, depending on the mediator mass~\cite{LZ:2022lsv}.

\underline{\it Indirect detection from galactic center emission}: If DM BSF  occurs in the Sun, it is reasonable to anticipate its occurrence in the galactic center of the Milky Way as well. Consequently, we would expect corresponding emission of  energetic neutrinos originating from the galactic center, also with a box-shaped energy spectrum. Here we estimate the indirect search limit by re-scaling the current bounds on symmetric DM from neutrino telescope observations.  Regarding indirect signals, a BSF process that generates one mediator and eventually two neutrinos with  $E_\nu \approx E_{\rm bind}/2$ is equivalent to symmetric DM annihilation with a mass of $E_{\rm bind}/2$,  if the latter only makes up a  fraction of the observed DM abundance. Therefore, existing bounds can be rescaled as follows, 
\begin{equation}\label{eq:IDrescale}
 (\sigma_{\rm symm.} v)|_{{m_\chi \to E_{\rm bind}/2}}  \rightarrow  \left({E_{\rm  bind}/2 \over m_\chi }\right)^2  ( \sigma_{\rm BSF} v )  \,, 
\end{equation}
where $(\sigma_{\rm symm.} v)$ denotes the known bounds on symmetric DM as a function of the DM mass.   %

Quantitatively, the neutrino flux generated from the BSF process of halo DM particles is estimated to be  
\begin{equation}
    \Phi^{\rm D}_{\nu} \sim 1 \,\text{cm$^{-2}$s$^{-1}$}  \left( {\text{GeV}\over m_\chi}\right)^2\left( {  \sigma_{\rm BSF} v \over 3\times 10^{- 22}\,{\rm cm}^3/{\rm s} } \right) \,.
\end{equation}
This is similar to boosted DM case~\cite{Agashe:2014yua}, which can vary mildly due to the uncertainty of the Galactic J-factor~\cite{Cirelli:2010xx}. 
The corresponding upper bound on $\alpha$ is shown in the right panel of Fig.~\ref{fig:indirectB}, obtained from re-scaling the indirect bounds holding in the symmetric DM case~\cite{Arguelles:2019ouk}. As also illustrated in Fig.~\ref{fig:SIDM}, this excludes values of $m_\chi$ between $3.0$\, and $ 11.4$~GeV for $\alpha=0.15$, and gives irrelevant bound for $\alpha=0.05$.

We can further speculate when this galactic neutrino flux is larger than the one induced by BSF inside the Sun. The latter is approximately $\Phi^\odot_{\nu}/({\rm GeV}/m_\chi)^2 \sim \mathcal O(1)$\,cm$^{-2}$s$^{-1}$ when the DM capture on DMBS is  saturated by its geometric rate.  Using Eq.~\eqref{BSFrate} results in  $\alpha^{5/2}/m_\chi \lesssim 10^{-4}/{\rm GeV}$, which in turn requires  $m_\chi \gtrsim 5.6\,$GeV  (87.1\,GeV) for $\alpha =0.05$ (0.15),  see also the right panel of Fig.~\ref{fig:indirectB}.  In practice, the non-vanishing mass of the mediator suppresses the BSF cross section,  reducing the neutrino flux,  but barely affects the BSF-induced flux from the Sun.

\underline{\it CMB}: Since the CMB observables are mostly sensitive to the total electromagnetic energy injected by extra processes in the high-redshift Universe, the associated constraint on BSF can be obtained by re-scaling the CMB bound on the annihilation of a symmetric DM candidate~\cite{Elor:2015bho},
\begin{equation}
f_{\rm eff} {\sigma_{\rm BSF} v \over m_\chi} \, \frac{{\rm Br}_{\phi \to \hbox{\tiny EM} } E_{\rm bind}}{2m_\chi}\lesssim 4.1\times 10^{-28}\,\frac{\hbox{cm}^3/\hbox{s}}{\hbox{GeV}}\,.
\label{CMBbound}
\end{equation}
Here, $f_{\rm eff}$ is an efficiency factor that depends on the spectrum of injected electrons and photons, and  we have taken into account that the energy injected per process is not about $2m_\chi$ as in the case of DM annihilation, but is given by the binding energy $\approx \alpha^2 m_\chi/4$, times the electromagnetic branching ratio of $\phi$ decay, ${\rm Br}_{\phi \to \text{\tiny EM}}$.  For $m_{\phi} \ll E_{\rm bind} $, the mass of the light mediator does not play any role in the CMB bound. For decay into SM particles other than neutrinos, the fact that BSF is strongly enhanced when the DM relative velocity is small gives a CMB bound which is much stronger than other bounds derived from local cosmic-ray observations such as Fermi-Lat and AMS experiments. For instance, taking the DM velocity at CMB to be a few km/s leads to  ${\rm Br}^{1/3}_{\phi \to \text{\tiny EM}}\alpha^{7/3}\lesssim 10^{-5}m_\chi/\text{GeV}$.  Combining this condition with Eq.~\eqref{eq:widthnu} gives the constraint $Y_\phi \theta_\nu^2/\theta_\phi \gtrsim  10^{2.2}\alpha^{7/2}(\text{GeV}/m_\chi)^{3/2}$ for $m_\phi$ below twice the muon mass. For typical seesaw values of the neutrino mixing angle, Eq.~(\ref{sinalpha}), this translates to the condition $m_N\lesssim 22\,\hbox{TeV}\cdot Y_\phi\, (m_\chi/5\,\hbox{GeV})^{3/2}(10^{-12}/\theta_\phi)(0.1/\alpha)^{7/2}(m_\nu/0.1\,\hbox{eV})$.

\underline{\it Mediator decay into right-handed neutrinos}: 
If $2 m_N<m_\phi$ the dominant decay is not anymore into a pair of light neutrinos but into a pair of right-handed neutrinos (or into $\nu N$ for $m_\phi/2<m_N<m_\phi$), with their decay widths given by,
\begin{align}
\Gamma(\phi\rightarrow NN)=  \frac{1}{16\pi} |Y_ \phi|^2 \, m_\phi \,,  &&
\Gamma(\phi\rightarrow N \nu)=    \frac{1}{16\pi} |Y_ \phi|^2  \,m_\phi \,\sin^2\theta_\nu  \,.      
\end{align}
Using the typical seesaw expectation of Eq.~(\ref{sinalpha}), one gets
\begin{align}
|Y_\phi|^2 \sim   10^{-22}  \, \,\frac{1\,\hbox{GeV}}{m_\phi} \frac{1\,\hbox{s}}{\tau_\phi}\,,   &&
 |Y_\phi|^2 \sim  10^{-12}  \, \,\frac{1\,\hbox{GeV}}{m_\phi} \frac{1\,\hbox{s}}{\tau_\phi} \,\frac{0.1 \hbox{eV}}{m_\nu} \frac{m_N}{1 \hbox{GeV}}    
\end{align}
for $2 m_N<m_\phi$ and $m_\phi/2<m_N<m_\phi$, respectively.  This can be compatible with the  extra radiation constraint $\tau_\phi \lesssim 1$\,s but is basically excluded by the CMB bound because the right-handed neutrino decay product contains a non-negligible amount of electromagnetic material.

%%%%%%%%%%%%%%%%%%%%%%%%%%%
\section{Summary}

We have considered the possibility that asymmetric DM forms bound-states in the Sun, and showed that this leads to novel phenomenology. BSF in the Sun can proceed via emission of light scalar particles that carry energy roughly equal to the binding energy. Their decays to neutrinos lead to potentially testable low energy signals at neutrino detectors.  

Unlike for annihilating DM, BSF produces a flux of particles without reducing the number of DM particles in the Sun. We point out that on top of the DM particles captured in the Sun, the bound-states piling up in this way become additional scattering targets through which DM from the galactic halo could be captured. We have determined the associated DM accretion rates on DM and DM bound-states by evaluating the differential cross section, taking into account that for typical parameters,  $v\sim 10^{-3}$, $\alpha \sim 0.1$, $m_\phi< $ GeV and $m_\chi \sim 100$ GeV, the DM-DM and DM-DMBS scattering processes proceed in the semi-classical regime. As soon as these self capture rates are larger than $t_\odot^{-1}$, they can become phenomenologically relevant. In particular, we have shown that, thanks to the self-capture on bound-states, the number of DM particles in the Sun can exponentially increase, so much that the capture rate can reach the geometric rate, i.e. all the DM particles intercepting the DM bound-state thermal sphere are captured as the mean free path becomes smaller than the sphere. As a result, this exponential effect also considerably boosts the BSF and thus the associated flux of light mediators. In an example model, where DM is a Dirac fermion which self-interacts through exchange of light scalar that mixes with the Higgs boson, with the scalar decaying into two neutrinos through seesaw interactions, this leads to a neutrino flux
which reach the predicted atmospheric neutrino fluxes at energies below hundred MeV. Near future experiments such as Hyper-K, as well as direct detection experiments, will be able to probe further this scenario.   

%%%%%%%%%%%%%%%%%%%%%%%%%
\paragraph*{{\bf Acknowledgments:}} We thank    Sergio Palomares-Ruiz and Hai-bo Yu for discussions, and Sebastian Wild for help with the implementation of direct detection rates in DDCalc~\cite{GAMBIT:2018eea}. 
X.C. is supported by the Research Network Quantum Aspects of Spacetime (TURIS)  and co-funded by the European Union (ERC, NLO-DM, 101044443). 
R.G. is supported by MIUR grants PRIN 2017FMJFMW  and 2017L5W2PT, and thanks Galileo Galilei Institute for hospitality. C.G.C. is supported by a Ramón y Cajal contract with Ref.~RYC2020-029248-I, the Spanish National Grant PID2022-137268NA-C55 and Generalitat Valenciana through the grant CIPROM/22/69. The work of T.H. is supported by the Excellence of Science (EoS) project No. 30820817 - be.h “The H boson gateway to physics beyond the Standard Model”, by the IISN convention 4.4503.15, and by CERN where part of this work has been done. The authors acknowledge the workshop on Self-Interacting Dark Matter: Models, Simulations and Signals 2023, Pollica, for hospitality.

%%%%%%%%%%%%%%%%%%%%%%%%%%%%%%%%%%%%%%%%%%%%%%%
%\newpage
\appendix
\section{Approximate parametric solution to number evolution of DM particles}\label{App:nx_evolve}

The set of coupled Boltzmann equations for $N_\chi$ and $N_{2\chi}$, Eqs.~\eqref{eq:boltz}-\eqref{eq:boltz2}, has no closed-form analytical solution. However, we can construct an approximate solution  and better understand the underlying physics by proceeding step by step as follows. 

\subsection{Case with no self-capture on DM bound-states}

Without the $C_{2\chi}$ term, the equation for $N_\chi$ does not depend on $N_{2\chi}$ 
 \bea
\label{eq:boltzA}
\frac{\dd N_{\chi}}{\dd t} &=& C_\star - A_{\rm bsf} N^2_\chi + C_\chi N_\chi\,, 
\ena
which can be solved easily as follows.

\paragraph*{ \underline{BSF with capture only on nucleons: $C_\chi= C_{2\chi} =0$ }:}
 Without any capture due to DM self-interactions, the above equation has the solution
\beq
N^0_{\chi}(t ) = C_\star \tau_0 \tanh\left(\frac{t}{\tau_0}\right)    \quad\quad, \quad \quad N^0_{2\chi}(t ) = {1\over 2}C_\star \left(t -  \tau_0 \tanh\left(\frac{t}{\tau_0}\right)  \right)~, 
\label{eq:Nchizero}
\eeq
with the time constant given by $\tau_0 = (C_\star A_{\rm bsf})^{-1/2}$. 
Without BSF, i.e. when $\tau_0\to\infty$, this solution gives a $N_\chi$  linearly increasing in time as expected, $N^0_{\chi}(t \lesssim \tau_0 ) = C_\star t$. With BSF, such an increase occurs until
there are enough DM particles for BSF to proceed. For $t\gtrsim \tau_0$, an equilibrium between BSF and capture on nucleons is reached so that $N_\chi$ saturates becoming constant: $N_\chi=C_\star \tau_0=(C_\star/A_{\rm bsf})^{1/2}$. This value can be obtained also directly from the Boltzmann equation, Eq.~(\ref{eq:boltz}), whose right-hand side vanishes when $C_\star-A_{\rm bsf}N_\chi^2=0$.
As expected, the larger $A_{\rm bsf}$, the sooner the saturation is reached and the smaller the asymptotic value of $N_\chi$ is. Conversely, the larger $C_\star$,  the larger  $N_\chi$ must be for the term associated with BSF to compensate that of $C_\star$. Note, nevertheless, that the larger $C_\star$ and the larger $A_{\rm bsf}$, the smaller the saturation time scale $\tau_0$. This stems from the fact that the number of particles captured increases more rapidly, and BSF starts to become important earlier. Consequently, BSF catches up with the capture process sooner.

\paragraph*{ \underline{BSF with capture on nucleons and on free DM particles: $C_\chi\neq 0,\, C_{2\chi} =0$ }:}
Early in the evolution the contribution of $C_{2\chi}  N_{2\chi}$ is negligible.  The analytical solution when $C_{2\chi}=0$ and $C_\chi\neq 0$ is 
\begin{eqnarray}
N^s_{\chi}(t ) &=& \frac{C_\chi}{2 A_{\rm bsf }} +\frac{1}{A_{\rm bsf }\tau_s} \tanh\left(\frac{t}{\tau_s} -{\rm arctanh}\left(\frac{C_\chi \tau_s}{2}\right)\right)
\,, \label{eq:sol_cs}\\
2N^s_{2\chi}(t)+ N^s_{\chi}(t )&=& \left(C_\star+\frac{C_\chi^2}{2 A_{\rm bsf }}+\frac{C_\chi}{ A_{\rm bsf } \tau_s}\right)t +\frac{C_\chi}{A_{\rm bsf }} \log\left(\frac{2-C_\chi \tau_s +e^{-\frac{2t}{\tau_s}} \left(2+C_\chi \tau_s\right) }{4}\right) \,,\label{eq:sol_n2x_cs} 
\end{eqnarray}
with $\tau_s = (C_\star A_{\rm bsf} +C^2_\chi/4)^{-1/2}$.   As it must be, for $C_\chi=0$ these equations reduce  to Eq.~\eqref{eq:Nchizero}  and $\tau_s$ reduces to $\tau_0$. 

Here again, the BSF term equilibrates with capture in Eq.~(\ref{eq:boltz}), as it is negative and  quadratic in $N_\chi$,  whereas the $C_\star$ ($C_\chi$) term is constant (linear) in it.
Thus, $N_\chi$ saturates when the right-hand side of Eq.~(\ref{eq:boltz}) vanishes, $C_\star-A_{\rm bsf}N_\chi^2+C_\chi N_\chi=0$, which gives  $N^{s,\rm eq}_\chi =C_\chi/(2 A_{\rm bsf}) +1/(\tau_s A_{\rm bsf}) =  C_\star/(\tau_s^{-1} - C_\chi/2 )$.
The equilibrium time scale $\tau_s$ is smaller than that without the $C_\chi$ term, $\tau_0$, because the $C_\chi$ term increases the capture, so that BSF becomes important earlier.
Note that before equilibrium is reached the solution of Eq.~(\ref{eq:sol_cs}) is exponential. However the effect of this exponential is very limited as the equilibrium  is reached soon, as can be seen in Fig.~(\ref{fig:number_evolution}), for various examples of parameter sets.\footnote{Actually the solution of Eq.~(\ref{eq:sol_cs}) can be rewritten in exponential form as  $2 C_\star \tau_s (e^{2t/\tau_s}-1)/ (A+B e^{2t/\tau_s})$ with $A=-2-C_\chi \tau_s$ and $B= -2+C_\chi \tau_S$. In practice if $C_\star A_{\rm bsf} < C^2_\chi/4$ (as will be the case for our scenario), then $\tau_s\simeq 2/C_\chi$ which gives $B<A$ so that $N_\chi\simeq 2 C_\star \tau_s (e^{2t/\tau_s}-1)/A$, which is exponentially growing until it reaches the equilibrium plateau when $B e^{2t/\tau_s}$ becomes larger than $A$. The amount of exponential grows is limited because in practice $B$ is not much smaller than $A$.} Note also that the total number of particles grows linearly, Eq.~(\ref{eq:sol_n2x_cs}), except for a logarithmic correction, particularly for $t\gg\tau_s$.  More importantly, for realistic values of  $C_\star$ and $C_\chi$, the rate associated with the saturation value always lies well below the geometric rate $C^g_\chi$ of thermalized DM particles in the Sun.

%%%%%%%%%%%%%%%%%%%%%%%%%%%%%%%%%%%%%%%%%%%%%%%%%%%%%%%
\subsection{Case with self-capture on DM bound-states}

The appearance of a non-vanishing $C_{2\chi}$ term drastically changes the physics  in several ways. First of all, the $C_{2\chi}$ term implies that the number of free DM particles in the Sun depends on the number of DM bound-states so that the Boltzmann equations for $N_\chi$ and $N_{2\chi}$ are coupled. 
Differentiating  Eq.~\eqref{eq:boltz} with respect to time and using the equation for the bound-state, Eq.~\eqref{eq:boltz2}, results in the following second-order differential equation,
\bea
\label{eq:self_boltz}
\frac{\dd^2 N_{\chi}}{\dd^2 t} &=& \left(-2A_{\rm bsf} N_\chi +C_\chi \right) \frac{\dd N_{\chi}}{\dd t} + \frac{1}{2}C_{2\chi} A_{\rm bsf} N^2_\chi\,,
\ena
with boundary conditions: $\frac{\dd N_{\chi}(t=0)}{\dd t}= C_\star$ and $N_\chi(t=0)=0$. Formally, Eq.~\eqref{eq:self_boltz} is of the form $ N^{\prime \prime} = f(N) N^\prime + g(N)$, known as a Li\'enard equation, with no known general analytical solutions. 
 
 Second, the self-capture associated with $C_{2\chi}$ is larger than that induced by $C_\chi$. This and the fact that $N_{2\chi}$ never saturates --as Eq.~\eqref{eq:boltz2} shows--  
imply that the  term  $C_{2\chi} N_{2\chi}$ in Eq.~(\ref{eq:boltz}) becomes much larger than $C_\chi N_\chi$. In fact,  $N_\chi$ saturates to values much larger than in the case of $C_\chi\neq 0, C_{2\chi}=0$, because the right-hand side of Eq.~(\ref{eq:boltz}) vanishes for  $A_{\rm bsf}N_\chi^2\simeq C_{2\chi} N_{2\chi}\gg C_\chi N_\chi$. Thus, unlike in the case of $C_\chi\neq 0,\,C_{2\chi}=0$, where the exponential growth induced by the $C_\chi$ term does not last long, the $C_{2\chi}$ term induces an exponential growth for $N_\chi$ which is both faster (since the argument of the exponential will be proportional to $C_{2\chi}$ rather than to $C_\chi$) and much larger. 
Therefore, $N_\chi$  can reach in this way much higher values, large enough for the capture rate to reach the geometric limit within the DMBS thermal sphere.  Approximate solutions can be obtained as follows.

\paragraph*{ \underline{BSF with capture on nucleons and on DM bound-states: $C_\chi= 0,\, C_{2\chi} \neq 0$ }:}
 
For the parameters of interest in this work, $C_{2\chi} \ll 100/\tau_0$, we  numerically find that the following exponential ansatz provides a good approximation from $t = \tau_0$ to the point when $C_{2\chi} N_{2\chi}$ saturates the geometric capture rate.  That is,  in practice
 \beq\label{eq:nx_sols_c2s_only}
 N_\chi(\tau_0 \lesssim t \lesssim \tau_g) = N_\chi^0(\tau_0) \exp\Bigg[{ C_{2\chi} \over 4}  (t-\tau_0)  \Bigg]~.
 \eeq
This is also suggested by the quasi-static solution of $dN_\chi/dt \sim 0$ in Eqs.~\eqref{eq:boltz} and \eqref{eq:boltz2}, where  
\begin{equation}\label{eq:expAppr}
\frac{dN_{2\chi}}{dt} \simeq \frac{C_\star}{2} + \frac{C_{2\chi} }{2} N_{2\chi} \,~~\text{and~~~} N_\chi \simeq  \left(\frac{ C_\star + C_{2\chi} N_{2\chi} }{  A_{\rm bsf}}\right)^{1/2}\,.  
\end{equation} 
For the opposite case, $C_{2\chi} \ge 100/\tau_0$, the divergence happens even faster.

Therefore, capture on bound-states exponentially increases the number of DM particles within the Solar lifetime if the condition, $t_\odot -\tau_0 \gg  C_{2\chi}^{-1}$ is satisfied.  The fact that the exponential growth starts when $N_{2\chi}$ becomes of order $N_\chi$ can also be seen in the numerical examples shown in Fig.~\ref{fig:number_evolution}.

%%%%%%%%%%%%%%%%%%%%%%%%%%%%%%%
\paragraph*{ \underline{BSF in the full general case: $C_\chi\neq 0,\, C_{2\chi} \neq 0$ }:} 
For the reasons explained above, switching on the $C_\chi$ term --in addition to the $C_{2\chi}$ term-- does not drastically change the result. It induces an additional moderate exponential growth that makes the contribution of  $C_{2\chi}$ important slightly earlier (i.e.~around $t=\tau_s$ rather than at $\tau_0$), see Fig.~\ref{fig:number_evolution}.
 An approximate solution to this general case is obtained in the same way
 as Eq.~(\ref{eq:nx_sols_c2s_only}), by matching Eq.~(\ref{eq:sol_cs}) at $\tau_s$ rather than at $\tau_0$. Interestingly,  since $\tau_s < 2/C_\chi \sim 4/C_{2\chi} $, we have $\tau_s  C_{2\chi} \lesssim 4$. That is, the condition for the validity of  Eq.~\eqref{eq:nx_sols_c2s_only}, $C_{2\chi} \ll 100/\tau_0$,  in which $\tau_0$ is now  replaced by $\tau_s$,  is automatically satisfied.  This suggests
\beq\label{eq:nx_sol_all}
N_\chi(\tau_s\lesssim t \lesssim \tau_g) =  N^s_\chi(\tau_s) \exp\Bigg[{{ C_{2\chi }\over 4}  \left(t-\tau_s\right)}\Bigg]~,
\eeq
where  $N^s_\chi(\tau_s) $ is well approximated by Eq.~\eqref{eq:sol_cs}. 

This exponential growth lasts until the geometric capture rate within the DMBS thermal sphere is saturated.  Quickly after $t= \tau_g$, $N_\chi$ stops increasing when the BSF term (quadratic in $N_\chi$) compensates the constant capture rate.\footnote{Once the capture rate reaches the DMBS sphere geometric rate, the time it takes for $N_\chi$ to become constant is small, $\Delta t\simeq 1/\sqrt{A_{\rm bsf} C^g_{2\chi}}$.} The quasi-static equilibrium solution,   obtained from $dN_\chi/dt\simeq C_\star+C_{2\chi}^g -A_{bsf} N_\chi^2\simeq 0$, yields the final particle number\footnote{Here we neglect the subleading capture contribution, from the $C_\chi$ term, within the DM thermal sphere outside the DMBS thermal sphere. 
 To be  accurate the $C_\star$ term in Eq.~(\ref{eq:posttau1Nchi}) has to be taken into account only outside the DMBS thermal sphere as the $C_{2\chi}^g$ term already accounts for the maximal total capture rate within this sphere, but this concerns a negligible effect.
}
 \beq\label{eq:posttau1Nchi}
 N_\chi(t>\tau_g) = \left(\frac{C^g_{2\chi} + C_\star}{A_{\rm bsf}}\right)^{1/2}\,.
 \eeq
The associated time $\tau_g$, where such an equilibrium is reached, is determined as
\beq
 \label{tau1}
\tau_g = \tau_s + \frac{2}{C_{2\chi} }\log \left(\frac{C^g_{2\chi} +C_\star}{A_{\rm bsf} (N^s_\chi(\tau_s))^2}\right) \,.
\eeq
The condition that $\tau_g \leq t_\odot$, so that the DM number can be maximized at present, is satisfied if
 \beq
\,\frac{C_{2\chi}}{2} \left(t_\odot - \tau_s \right) \geq \log \left(\frac{C^g_{2\chi} +C_\star}{A_{\rm bsf} (N^s_\chi(\tau_s))^2}\right)  = \begin{cases}
      \log \left(\frac{ C^g_{2\chi} /  C_\star +1  }{\tanh[1]^2} \right)  & \text{if~} C_\chi^2  \le A_{\rm bsf} C_\star\\
      \log \left(\frac{ C^g_{2\chi} /  C_\star +1  }{(e^2-1)^2 }\, \frac{ C_\chi^2  }{A_{\rm bsf} C_\star } \right)  & \text{if~} C_\chi^2  \gg A_{\rm bsf} C_\star
    \end{cases} \, .
\label{tau1versustauSun}
 \eeq
The latter case has $\tau_0 \gg \tau_s \simeq 2/C_{\chi} \sim 4/C_{2\chi}$, so the saturation should happen within a few $\tau_s$. 

We can now analytically match the above two solutions at times $\sim\tau_s$ and $\tau_g$. 
The final approximate solution for species $i$ has the parametric form
\beq\label{eq:sol_full}
N_i(t) ~ = ~ N_i^{t<\tau_s}  \Theta(\tau_s -t) + N_i^{\tau_s<t<\tau_g}  \Theta(t- \tau_s ) \Theta(\tau_g -t ) + N_i^{t>\tau_g}  \Theta(t -\tau_g) \,,
\eeq
where  $N_i^{t<\tau_s}$, $N_i^{\tau_s<t<\tau_g}$ and $N_i^{t>\tau_g}$ are given by Eqs.~\eqref{eq:sol_cs}, ~\eqref{eq:nx_sol_all} and ~\eqref{eq:posttau1Nchi}, respectively.
We have checked that these analytical expressions agree well with our numerical results presented in the various figures from solving the Boltzmann equations. 

%%%%%%%%%%%%%%%%%%%%%%%%%%%%%%%%%%%%%%%%%%%%
%%%%%%%%%%%%%%%%%%%%%%%%%%%%%%%%%%%%%%

\section{(Non-)perturbative treatment of elastic DM scattering}
\label{app:nonpertscatt}

Suppose that DM scatters off a target with a mass $m_T$ with a negligible form factor, the cross section differential in the recoil energy is 
\begin{align}
\frac{\dd \sigma}{\dd E_R} = \frac{2\pi\,m_{T}}{k^2} \big|{\cal M}(k,\theta)\big|^2\,,&&\text{with} && \cos\theta =1 - \frac{m_{T} E_R}{k^2}\,.% \cos\theta =1 - \frac{4 E_R}{m_\chi w^2},
\label{eq:dsigmadErApp}
\end{align}
Here $\theta$ and $k=\mu v$   are respectively the scattering angle and the incoming momentum in the center of mass frame, while the reduced mass of the DM-target system $\mu = m_\chi m_T/(m_\chi +  m_T)$.  We aim to calculate this when the scattering is triggered by the exchange of a scalar mediator, particularly when non-perturbative effects cannot be disregarded. In the non-relativistic limit, this is described by a free DM particle scattered by the corresponding Yukawa potential
$
V(r)= -\alpha' e^{-m_\phi r}/r$, with $\alpha'=
      g_s y_N \sin\theta_\phi /4\pi, \alpha $ or $2\alpha$ when target is a nucleon, a DM particle or a DMBS, respectively.

The amplitude of such quantum scattering can be  obtained from a partial-wave expansion  
\begin{align}
 {{\cal M}}(k,\theta)&=\frac{1}{2ik}\sum^\infty_{\ell=0} (2l+1)  P_\ell \left(\cos \theta\right)\left(e^{2i \delta_\ell}-1\right)\,,
\label{eq:Mdef}
\end{align}
where $\ell$ is the orbital angular momentum, $P_\ell(\cos\theta)$ is the corresponding Legendre polynomial and $\delta_\ell$ is the phase-shift. 
In general, the phase-shift must be obtained by solving the radial part of the Schr\"odinger equation describing the collision, which is equivalent to solving~\cite{Chu:2019awd} 
\begin{align}
\delta_{\ell,k} '(r)= -2k\mu \, r^2 V(r) {\text{ Re}}\left[ e^{i \delta_{\ell,k}(r) }h^{(1)}_{\ell}(kr)\right]^2\,
\,  \text{~with~}
\delta_{\ell,k} (0) = 0\, \text{~and~~} \delta_{\ell,k}(r)|_{r\to\infty}\to \delta_\ell \,.
\label{eq:phaseshifts}
\end{align} 
Here,  $h^{(1)}_{\ell} =j_\ell+i \,n_\ell $ is the spherical Hankel function of the first kind.  Nevertheless, solving  Eq.~\eqref{eq:phaseshifts} for all values of $\ell$ is impractical, and thus simplifications are necessary for different parameter regimes, as shown in left panel of  Fig.~\ref{fig:regimes}. This will be discussed below for integrated and differential DM-DM(BS) scattering, separately. The Born approximation is always justified for the very weak scattering between DM and nucleon.

%%%%%%%%%%%%%%%%%%%%%%%%%%%%%%%%%%%%%%%%%%%%%%
\subsection{Integrated DM self-scattering in DM halos}
\label{app:selfscatt}

\begin{figure}[t]
\includegraphics[width=0.5\textwidth]{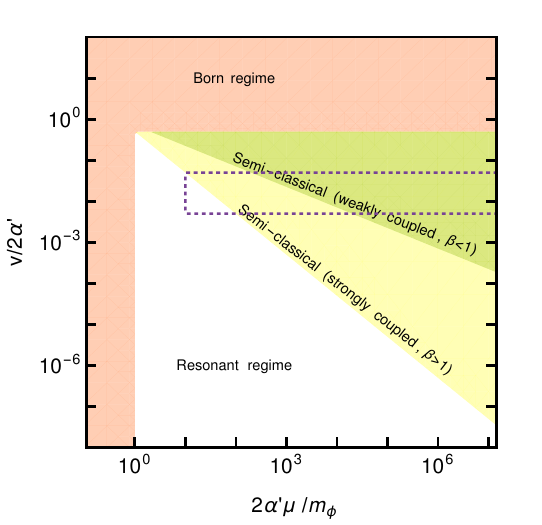}   
\includegraphics[width=0.49\textwidth]{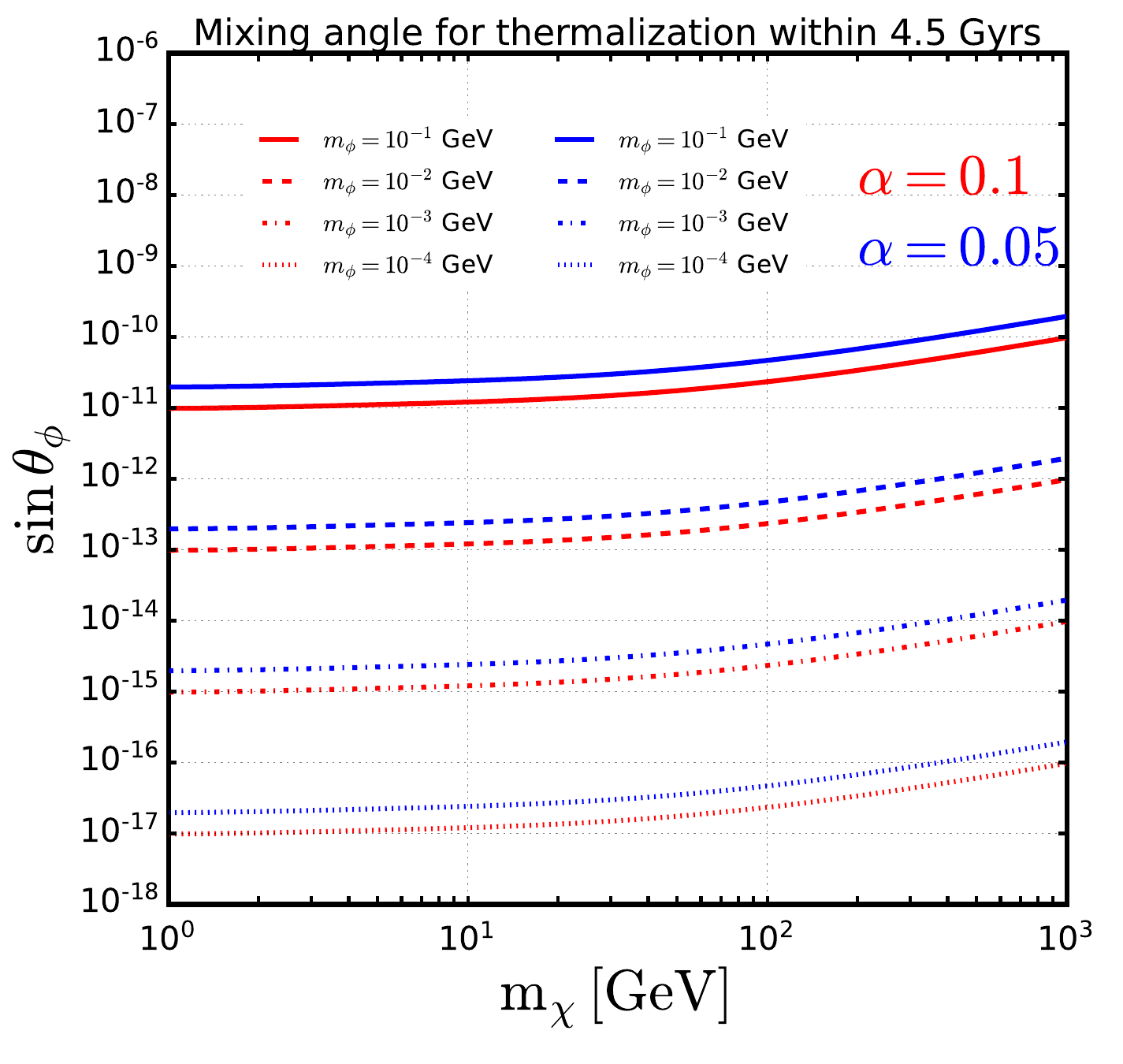}
	\caption{\emph{Left}: Dark matter self-scattering regimes.  
For self-capture of dark matter by the Sun, the relevant parameters  roughly lie within the dotted line, and correspond to $v\sim 4\cdot 10^{-3}$, $m_\chi \gtrsim 5$ GeV, $ 10^{-4}$ GeV $\lesssim m_\phi \lesssim 10^{-2}$ GeV, and $0.02  \lesssim \alpha < 0.2 $. See text for details. \emph{Right}: Curves above which the captured DM particles thermalize with the nucleons in the Sun within $t_\odot$.} 
\label{fig:regimes}
\end{figure}

Sufficiently large DM self-scattering can alter the evolution of   halos, which gives rise to constraints on the corresponding scattering rates.   As argued in Ref.~\cite{Yang:2022hkm}, the gravothermal evolution of such  halos is best
characterized by the so-called viscosity cross section, $ \sigma_V \equiv {3\over 2 }\int^{1}_{-1} d\cos \theta\, \sin^2\theta {d\sigma \over d\cos\theta }$.\footnote{Following Ref.~\cite{Yang:2022hkm}, we introduce a prefactor of $3/2$  to make  $\sigma_V = \sigma$ for isotropic scatterings.}  The parameter regimes where  the Born and semi-classical approximations apply are given in left panel of  Fig.~\ref{fig:regimes}.  

For $\alpha \lesssim v $ or $ m_\chi \lesssim m_\phi/\alpha $, the first-order Born approximation is justified~\cite{Landau:1991wop}. Note that $\mu=m_\chi/2$ and $\alpha'=\alpha$ in this case. Moreover, accounting for the indistinguishability of the DM particles in  $\chi -\chi$ scattering,  we find
\begin{equation}
   \sigma_V =  \frac{6 \pi  \beta ^2}{ m_{\phi
   }^2} \left( \frac{ 
   \left(\frac{m_\chi v}{m_\phi}\right)^4+ 5  \left(\frac{m_\chi v}{m_\phi}\right)^2 +5     }{   \left(\frac{m_\chi v}{m_\phi}\right)^4+2  \left(\frac{m_\chi v}{m_\phi}\right)^2 }  \log \left[ \left(\frac{m_\chi v}{m_\phi}\right)^2+1 \right] - \frac{ 5 }{2}\right) \,.
\end{equation}
Where we have introduced $\beta = 2 m_\phi \alpha/(m_\chi v^2)$ for later convenience.  Note that if one adopts the modified transfer cross section, defined as $\sigma_{|T|}  \equiv \int^{1}_{-1} d\cos \theta\,  (1-|\cos\theta|) {d\sigma \over d\cos\theta } $, approximately $\sigma_V \simeq  2\sigma_{|T|} $ if $ m_\chi v /m_\phi \ll 1$ as well as  $\sigma_V \simeq  3 \sigma_{|T|} $ if $ m_\chi v /m_\phi  \gg 1$.
 
Outside the Born regime,  non-perturbative effects are non-negligible, especially in the resonant regime ($\alpha m_\chi  /m_\phi \gtrsim 1$ and $m_\chi v/m_\phi  \lesssim 1$). 
In spite of this, in this parameter region the scattering is generally dominated by the $s$-wave, $\ell=0$~\cite{Tulin:2013teo}, for which we can adopt the analytical results obtained by solving for the phase shift, $\delta_0$ of $\ell=0$,  under the assumption that the Yukawa potential can be approximated by a Hulth{\'e}n potential~\cite{Cassel:2009wt}. 
For identical fermions, after adding the symmetric factor $1/2$ for final states, this leads to 
\begin{equation}
 \sigma_V = {4\pi\over 2 }  { d\sigma \over d\Omega} \bigg|_{\ell=0}   =   {2\pi \over   k^2 } \Bigg|e^{i\delta_0} \Bigl( P_0 (\cos\theta ) +P_0 (-\cos\theta ) \Bigr) \sin\delta_0 \Bigg|^2  =  {8\pi \over  k^2 } \Bigg|e^{i\delta_0}    \sin\delta_0 \Bigg|^2  \,,
\end{equation}
where the scattering is isotropic, $k= m_\chi v/2$,  and $P_0$ is the $0$th-order Legendre polynomial. %, leading to a.  
Note that this cross section only includes an even spatial wave-function and an anti-symmetric spin wave-function.  We take the probability of such anti-parallel spin alignment to be 1/4. That is, its corresponding cross section  for identical fermion is reduced by half  with respect to that of non-identical fermions.  The characteristic peaks  of this regime are visible in Figs.~\ref{fig:rates_cs} and \ref{fig:SIDM}.

Finally, with the resulting expressions of the viscosity cross section for all regimes, we impose bounds on DM self-interaction at both cluster and dwarf scales, obtaining the observationally allowed parameter region shown in Fig.~\ref{fig:SIDM}, see also section~\ref{sec:bounds}.

%%%%%%%%%%%%%%%%%%%%%%%%%%%%%%%%%%%%%%%%%%%%%%%%%%
%%%%%%%%%%%%%%%%%%%%%%%%%%%%%%%%%%%%%%%%%%%%%%%%%%
\subsection{Differential DM-DM(BS) scattering for DM capture}
\label{app:selfcap}

As explained in the main text and above, while the impact of DM self-interactions in galaxies and galaxy clusters can be characterized by the integrated transfer or viscosity total cross sections, for DM self-capture in the Sun, it is crucial to obtain  the differential scattering cross section.  
In the latter case, left panel of Fig.~\ref{fig:regimes} (dotted contour) illustrates the parameter region of interest for the DM capture via scatterings off free and bound DM particles already captured in the Sun.  

The relevant parameters largely lie in the semi-classical regime, where $k\gg m^{-1}_{\phi}$.  That is, the de Broglie wavelength is much larger than the range of the Yukawa potential and the scattering amplitude can be estimated semi-classically. 
More precisely, the resulting sum in Eq.~\eqref{eq:Mdef}  can be performed using the stationary phase approximation~\cite{Landau:1991wop}, which, up to an inconsequential  global phase, gives
\begin{align}
{\cal M}= \frac{1}{k}\left(\frac{\ell_0(\theta)}{\sin \theta} \Big| \frac{\dd\ell_0(\theta)}{\dd\theta}\Big|\right)^{\frac{1}{2}} 
\,,
&&
\text{where}
&&
\int^\infty_{r_0}\frac{\ell_0(\theta)+\frac{1}{2}}{r^2\sqrt{k^2- \left(\frac{\ell_0(\theta)+\frac{1}{2}}{r}\right)^2 - 2\mu V(r)}} \dd r   =  \frac{\pi+\theta}{2}\,.
\label{eq:classicalApp}
\end{align}
Here $r_0$ is the closest distance from the potential centre, and $\ell_0(\theta)$ is the angular-momentum minimizing the phase in Eq.~\eqref{eq:Mdef} for a given scattering angle $\theta$. All this applies as long as $\ell_0(\theta)\gg 1$, which justifies neglecting $1/2$ in Eq.~\eqref{eq:classicalApp}.  To see how Eq.~\eqref{eq:classicalApp}  resembles the classical equations of motion, we define the impact parameter as $\rho \equiv \ell_0(\theta)/k$, which together with Eq.~\eqref{eq:dsigmadErApp}, leads to
\begin{align}
\frac{\dd \sigma}{\dd \cos\theta} =  2\pi \rho  \frac{\dd\rho}{\dd\cos \theta}
\,,
&&
\text{where}
&&
\int^\infty_{r_0}\frac{\rho \,\dd r}{r^2\sqrt{1- \frac{\rho^2}{r^2} + 2 \beta \left(\frac{e^{-m_\phi r }}{ m_\phi r} \right)}}    =  \frac{\pi+\theta}{2}\,,
\label{eq:classicalAppN}
\end{align}
with $\beta \equiv  \mu m_\phi \alpha'/k^2 $.  The integral in Eq.~\eqref{eq:classicalAppN}  gives the scattering angle according to classical mechanics. In this work, we follow the numerical approach presented in Ref.~\cite{Khrapak:1308514, Khrapak:2003kjw} to solve Eq.~\eqref{eq:classicalAppN} and obtain the scattering rates.

As shown by Ref.~\cite{ Khrapak:1308514}, the calculation can be further simplified for $\beta>13.2$, where the angular dependence of ${\dd \sigma}/{\dd \cos\theta}$ becomes insensitive to $\beta$. Performing a numerical fit to the cross sections of non-identical scattering and defining ${\dd \sigma}/{\dd \cos\theta} \equiv  \pi \rho^2_{*}(\beta,m_\phi) {\mathcal N}(\cos\theta)$, we find
\begin{equation}
\label{eq:fitangular}
\mathcal{N}(c_\theta) \simeq 0.35 + 0.05 c_\theta + 2.09 c_\theta^3 + 3.25 c_\theta^4 -5.51 c_\theta^5 -  8.53 c_\theta^6 + 5.06 c_\theta^7 + 7.22 c_\theta^8 \,.  
\end{equation}
Note that the normalization, $\int^1_{-1}   dc_\theta (1-c_\theta) \mathcal{N}(c_\theta) \simeq 1$,  is chosen in a way  we can directly adopt  the overall prefactor from  \cite{Khrapak:2003kjw} as 
\begin{equation}
\label{eq:rho_classical}
  \rho^2_\ast(\beta, m_\phi)=
    \begin{cases}
      \frac{1}{m^2_\phi} \left( \log \beta +1 - \frac{1}{2\, \log \beta}\right)^2 &{\rm for} \quad \quad \beta >1000~,\\
       \frac{8}{m^2_\phi} \left( \frac{\beta^2}{1 + 1.5\, \beta^{1.65}}\right) & {\rm for}  \quad \quad1000> \beta > 13.2~.\\
    \end{cases}       
\end{equation}
The amplitudes at the vicinity of  $ \cos\theta = 1$, which may approach infinity, are not included in the numerical fit above, given that small-angle scatterings barely contribute to the self-capture rate as a result of $\theta_{\rm min}$ in Eqs.~\eqref{eq:cap-xx-self} and \eqref{eq:cap-x-2x-self}. 

Besides, under the assumption that  the interference is subleading in the (semi-)classical regime~\cite{landau1976mechanics}, the previous approach can also be applied to the scattering of identical particles by symmetrizing the differential scattering amplitude $\theta \to \pi -\theta$, as
\begin{equation}
    \frac{d\sigma}{d\cos\theta}\bigg|_{\rm \chi \chi} =  \frac{d\sigma}{d\cos\theta} + \frac{d\sigma}{d\cos\theta}\bigg|_{\cos\theta\to -\cos\theta}\,, 
\end{equation}
where the range of the scattering angle $\theta$ reduces to $[0, \,\pi/2]$.

%%%%%%%%%%%%%%%%%%%%%%%%%%%%%%%%%%%%%%%%%%%

\section{DM thermalization}
\label{ap:DMthermalization}
We summarize below the general formalism for DM thermalization with the Solar medium, and present the results applicable to the scenario presented in the main text. 
After being captured, DM particle orbits become smaller than the size of the celestial body. For contact interactions, the time taken for the orbits to shrink to its thermal radius is given by~\cite{Kouvaris:2010jy}
\beq
t_{\rm therm} = \frac{m_\chi}{\left(\sum_N \rho_N \sigma_{\chi N}\right)}\sqrt{\frac{m_\chi}{2\,T}}~. 
\eeq
For the Sun, the time scale is estimated via 
\bea
t^\odot_2 &=& 1.5\, {\rm yrs}\left(\frac{m_\chi}{10 \,\rm GeV} \right)^{3/2} \left(\frac{2.2\, \rm KeV}{T_c} \right)^{1/2} \left(\frac{150 \,\rm g/cm^3}{\rho_c} \right) \left(\frac{10^{-40} \, \rm cm^2}{\sigma_{\chi N}} \right)~.
\ena
Thermalization time is dominated by the last stages of the process with the typical energy and momentum transfer of the order of temperature~\cite{Tinyakov:2021lnt}. 
Formally, thermalization time is computed through the energy loss rate. Consider the elastic scattering of a DM particle $\chi$ off a distribution of target particles $T$: 
$\chi (k) + T (p) \rightarrow \chi (k^\prime) + T (p^\prime)$.
The interaction rate per DM particle reads~\cite{Bertoni:2013bsa,Garani:2020wge}:
%%%%
\bea
\label{eq:rate}
&&d \Gamma =2 \frac{\dd^3 k^\prime}{(2 \pi)^3} S(q_0,q)~, \nonumber\\
 &&S(q_0,q)=\iint \frac{\dd^3 p^\prime \dd^3 p }{(2 \pi)^6 2 E_{p^\prime} 2 E_{k^\prime}   2 E_{p} 2 E_k}  (2 \pi)^4 \delta^4\left(k+p -k^\prime -p^\prime \right) |\mathcal{M}|^2 f(E_p) \left(1 -f(E_{p^\prime})  \right)\,,
\ena
%%%
where the second line above is the response function, and $f(E_{p,p'})$ being the Maxwell-Boltzmann distributions w.r.t. the core temperature $T_c$. 
For non-degenerate medium, there exists little Pauli-blocking from the final states, so $1 -f(E_{p^\prime}) \simeq 1$. 
In turn, the rate of energy loss is given by
%%%
\beq
\Phi = 2 \int  \frac{\dd^3 k^\prime}{(2 \pi)^3} S(q_0,q)  \times (E_i - E_f)\,,
\label{eq:Elossrate}
\eeq
%%%
where  $k'$ is integrated from 0 to $k$. 

Using this we can write down the time taken to thermalize with the celestial body, i.e. to reach a final energy $E_f = 3/2\,T_{\rm c}$ starting with an initial kinetic energy $E_i =   m_\chi v_{\rm esc}^2/2$:
%%%%
\bea
\label{eq:ttime}
\tau_{\rm therm} = - \int^{E_f}_{E_i} \frac{\dd E_i}{\Phi}.
\ena
Integrating Eq.~\eqref{eq:rate} over $p'$, the response function becomes
\begin{eqnarray}
S(q_0,q)&=&\int \frac{\dd^3 p}{(2 \pi)^2 2 } \frac{|\mathcal{M}|^2} { 16 m_N^2  m^2_\chi} \delta^3  \delta\left(q_0 - E_p + E_{p^\prime}\right)  f(E_p), 
\label{eq:Sred}
\end{eqnarray}
where  $E_k = E_{k^\prime}  = m_\chi$ and  $E_p = E_{p^\prime} = m_N$ have been taken in the denominator, while for the numerator we use $E_p = m_N + p^2/2 m_N$ and $E_p^\prime = m_N + p^{\prime\, 2}/2 m_N$. 
Moreover, the $\delta^3$ enforces momentum conservation, $\mathbf{q} = \mathbf{p^\prime} -\mathbf{p}$, and the energy delta-function is recast in terms of angle between the incoming nuclei and the momentum transfer as shown in Ref.~\cite{Reddy:1997yr}, yielding
\beq
S(q_0,q) \approx \int \frac{\dd p \, \dd \cos \theta}{(2 \pi)^2 2 } 2\pi p^2 \frac{|\mathcal{M}|^2} { 16 m^2_N  m^2_\chi}  e^{\bar{\mu}/T_c} e^{-p^2/2 m_N T_c} \frac{m_N}{|p| |q|}\delta\left(\cos \theta -\cos \theta_0\right) \Theta(p^2- p^2_-)~, 
\label{eq:Smaxwell}
\eeq
along with
\beq
\cos \theta_0 = \frac{m_N}{|p||q|} \left(q_0 - \frac{q^2}{2 m_N}\right) \quad \quad {\rm and} \quad \quad p^2_- = \frac{m^2_N}{q^2} \left(q_0 - \frac{q^2}{2 m_N}\right)^2 ~.
\eeq
Integrating the above equation we get
\beq
S(q_0,q) \approx \frac{|\mathcal{M}|^2} { 64 \pi m^2_\chi} \frac{T_c}{|q|} e^{\bar{\mu}/T_c} e^{-p^2_-/2 m_N T_c}\,.
\eeq
Taking the Higgs-mixing model introduced in the main text, the squared amplitude for scattering on nucleon in the non-relativistic limit reads
\beq
|\mathcal{M}|^2 = g^2_s \cos^2_{\theta_\phi} \sin^2_{\theta_\phi}  \frac{m^2_N f^2_N}{v^2_H}\frac{m^2_\chi m^2_N}{\left(q_0^2 - q^2 - m^2_\phi\right)^2}~,
\eeq
which, in the limit of small energy transfer, corresponds to DM-nuclei scattering  cross section of 
\begin{equation}
    \sigma_{\chi N} =  g^2 _s \cos^2_{\theta_\phi} \sin^2_{\theta_\phi}  \frac{1}{16\pi (m_\chi +m_N)^2}\frac{m^2_N f^2_N}{v^2_H}\frac{m^2_\chi m^2_N}{m^4_\phi}\,.
\end{equation}

The $k^\prime$ integral is recast in terms of momentum transfer and energy transfer. The limits of integration are $ 0 <q_0< E_k $ and $q_0<q< 2 E_k - q_0$. Therefore the energy loss rate for a particle with initial momenta $k$ is
\beq
\frac{d \Gamma}{d q_0} \Delta E  =  \int \frac{ d q}{(2 \pi)^2} q \frac{E_{k^\prime}}{E_k} S(q_0,q) (E_k - E_{k^\prime}) = \int \frac{ d q}{(2 \pi)^2} q \frac{E_k -q_0}{E_k} S(q_0,q) q_0 ~.
\eeq
In the non-degenerate limit the above equation factorizes. Hence the total rate is proportional to cross section on one target particle times their number density~\cite{Reddy:1997yr}.  The corresponding numerical results are shown in the right panel of  Fig.~\ref{fig:regimes}, showing that the thermalization can be reached within the lifetime of the Sun, even with very tiny mixing angles.  Including the DM self-interactions would further reduce the time scale of thermalization.

\bibliographystyle{apsrev4-1}
\bibliography{refs}

\end{document}